\documentclass[twocolumn,letterpaper,aps,prc,superscriptaddress,showpacs,floatfix]{revtex4}

\usepackage{graphicx}	% Include figure filed

\usepackage{xspace}	% Include xspace

\newcommand{\pt}{\mbox{$p_T$}\xspace}

\newcommand{\raa}{\mbox{$R_{\rm AA}$}\xspace}

\newcommand{\Ncoll}{\mbox{$N_{\rm coll}$}\xspace}

\newcommand{\sqs}{\mbox{$\sqrt{s}$}\xspace}
\newcommand{\sqsn}{\mbox{$\sqrt{s_{_{NN}}}$}\xspace}

\newcommand{\rdau}{\mbox{$R_{d{\rm Au}}$}\xspace}
\newcommand{\rcp}{\mbox{$R_{\rm CP}$}\xspace}
\newcommand{\pp}{\mbox{$p$+$p$}\xspace}

\newcommand{\dau}{\mbox{$d+$Au}\xspace}  % specific nuclei are nonitalic
\newcommand{\ncol}[1]{\mbox{$\left< N_{\rm coll}(#1)\right>$}\xspace}

\newcommand{\midn}{\mbox{$\left|\eta\right|<0.35$}\xspace}
\newcommand{\midy}{\mbox{$\left|y\right|<0.35$}\xspace}
\newcommand{\fory}{\mbox{$1.2<y<2.2$}\xspace}
\newcommand{\forn}{\mbox{$1.2<\eta<2.4$}\xspace}
\newcommand{\bacy}{\mbox{$-2.2<y<-1.2$}\xspace}
\newcommand{\bacn}{\mbox{$-2.2<\eta<-1.2$}\xspace}
\newcommand{\muony}{\mbox{$1.2<|y|<2.2$}\xspace}

\newcommand{\jpsi}{\mbox{$J/\psi$}\xspace}  % J/\psi is fully italicized in PDB
\newcommand{\psip}{\mbox{$\psi'$}\xspace}

\newcommand{\mptsq}{\mbox{$\langle p_T^2 \rangle$}\xspace}
\newcommand{\mptsqmax}{\mbox{$\langle p_T^2 \rangle|_{p_T\leq p_T^{\rm max}}$}\xspace}

\newcommand{\mncol}{\mbox{$\left< N_{\rm coll}\right>$}\xspace}
\newcommand{\mevc}{\mbox{MeV/$c~$}}
\newcommand{\gevc}{\mbox{GeV/$c~$}}
\newcommand{\gevcsq}{\mbox{GeV/$c^{2}~$}}

\begin{document}


\title{Transverse-Momentum Dependence of the $J/\psi$ Nuclear Modification 
in $d+$Au Collisions at $\sqrt{s_{_{NN}}}$ = 200 GeV}

\newcommand{\abilene}{Abilene Christian University, Abilene, Texas 79699, USA}
\newcommand{\banaras}{Department of Physics, Banaras Hindu University, Varanasi 221005, India}
\newcommand{\barc}{Bhabha Atomic Research Centre, Bombay 400 085, India}
\newcommand{\bnlcoll}{Collider-Accelerator Department, Brookhaven National Laboratory, Upton, New York 11973-5000, USA}
\newcommand{\bnlphys}{Physics Department, Brookhaven National Laboratory, Upton, New York 11973-5000, USA}
\newcommand{\caucr}{University of California - Riverside, Riverside, California 92521, USA}
\newcommand{\charlesczech}{Charles University, Ovocn\'{y} trh 5, Praha 1, 116 36, Prague, Czech Republic}
\newcommand{\chonbuk}{Chonbuk National University, Jeonju, 561-756, Korea}
\newcommand{\ciae}{Science and Technology on Nuclear Data Laboratory, China Institute of Atomic Energy, Beijing 102413, P.~R.~China}
\newcommand{\cns}{Center for Nuclear Study, Graduate School of Science, University of Tokyo, 7-3-1 Hongo, Bunkyo, Tokyo 113-0033, Japan}
\newcommand{\colorado}{University of Colorado, Boulder, Colorado 80309, USA}
\newcommand{\columbia}{Columbia University, New York, New York 10027 and Nevis Laboratories, Irvington, New York 10533, USA}
\newcommand{\czechtech}{Czech Technical University, Zikova 4, 166 36 Prague 6, Czech Republic}
\newcommand{\dapnia}{Dapnia, CEA Saclay, F-91191, Gif-sur-Yvette, France}
\newcommand{\elte}{ELTE, E{\"o}tv{\"o}s Lor{\'a}nd University, H - 1117 Budapest, P{\'a}zm{\'a}ny P. s. 1/A, Hungary}
\newcommand{\ewha}{Ewha Womans University, Seoul 120-750, Korea}
\newcommand{\fit}{Florida Institute of Technology, Melbourne, Florida 32901, USA}
\newcommand{\fsu}{Florida State University, Tallahassee, Florida 32306, USA}
\newcommand{\gsu}{Georgia State University, Atlanta, Georgia 30303, USA}
\newcommand{\hiroshima}{Hiroshima University, Kagamiyama, Higashi-Hiroshima 739-8526, Japan}
\newcommand{\ihepprot}{IHEP Protvino, State Research Center of Russian Federation, Institute for High Energy Physics, Protvino, 142281, Russia}
\newcommand{\illuiuc}{University of Illinois at Urbana-Champaign, Urbana, Illinois 61801, USA}
\newcommand{\inrras}{Institute for Nuclear Research of the Russian Academy of Sciences, prospekt 60-letiya Oktyabrya 7a, Moscow 117312, Russia}
\newcommand{\instpasczech}{Institute of Physics, Academy of Sciences of the Czech Republic, Na Slovance 2, 182 21 Prague 8, Czech Republic}
\newcommand{\isu}{Iowa State University, Ames, Iowa 50011, USA}
\newcommand{\jinrdubna}{Joint Institute for Nuclear Research, 141980 Dubna, Moscow Region, Russia}
\newcommand{\jyvaskyla}{Helsinki Institute of Physics and University of Jyv{\"a}skyl{\"a}, P.O.Box 35, FI-40014 Jyv{\"a}skyl{\"a}, Finland}
\newcommand{\kek}{KEK, High Energy Accelerator Research Organization, Tsukuba, Ibaraki 305-0801, Japan}
\newcommand{\korea}{Korea University, Seoul, 136-701, Korea}
\newcommand{\kurchatov}{Russian Research Center ``Kurchatov Institute", Moscow, 123098 Russia}
\newcommand{\kyoto}{Kyoto University, Kyoto 606-8502, Japan}
\newcommand{\labllr}{Laboratoire Leprince-Ringuet, Ecole Polytechnique, CNRS-IN2P3, Route de Saclay, F-91128, Palaiseau, France}
\newcommand{\lawllnl}{Lawrence Livermore National Laboratory, Livermore, California 94550, USA}
\newcommand{\losalamos}{Los Alamos National Laboratory, Los Alamos, New Mexico 87545, USA}
\newcommand{\lpc}{LPC, Universit{\'e} Blaise Pascal, CNRS-IN2P3, Clermont-Fd, 63177 Aubiere Cedex, France}
\newcommand{\lund}{Department of Physics, Lund University, Box 118, SE-221 00 Lund, Sweden}
\newcommand{\maryland}{University of Maryland, College Park, Maryland 20742, USA}
\newcommand{\mass}{Department of Physics, University of Massachusetts, Amherst, Massachusetts 01003-9337, USA }
\newcommand{\muenster}{Institut fur Kernphysik, University of Muenster, D-48149 Muenster, Germany}
\newcommand{\muhlenberg}{Muhlenberg College, Allentown, Pennsylvania 18104-5586, USA}
\newcommand{\myongji}{Myongji University, Yongin, Kyonggido 449-728, Korea}
\newcommand{\nagasaki}{Nagasaki Institute of Applied Science, Nagasaki-shi, Nagasaki 851-0193, Japan}
\newcommand{\newmex}{University of New Mexico, Albuquerque, New Mexico 87131, USA }
\newcommand{\nmsu}{New Mexico State University, Las Cruces, New Mexico 88003, USA}
\newcommand{\ornl}{Oak Ridge National Laboratory, Oak Ridge, Tennessee 37831, USA}
\newcommand{\orsay}{IPN-Orsay, Universite Paris Sud, CNRS-IN2P3, BP1, F-91406, Orsay, France}
\newcommand{\peking}{Peking University, Beijing 100871, P.~R.~China}
\newcommand{\pnpi}{PNPI, Petersburg Nuclear Physics Institute, Gatchina, Leningrad region, 188300, Russia}
\newcommand{\riken}{RIKEN Nishina Center for Accelerator-Based Science, Wako, Saitama 351-0198, Japan}
\newcommand{\rikjrbrc}{RIKEN BNL Research Center, Brookhaven National Laboratory, Upton, New York 11973-5000, USA}
\newcommand{\rikkyo}{Physics Department, Rikkyo University, 3-34-1 Nishi-Ikebukuro, Toshima, Tokyo 171-8501, Japan}
\newcommand{\saispbstu}{Saint Petersburg State Polytechnic University, St. Petersburg, 195251 Russia}
\newcommand{\saopaulo}{Universidade de S{\~a}o Paulo, Instituto de F\'{\i}sica, Caixa Postal 66318, S{\~a}o Paulo CEP05315-970, Brazil}
\newcommand{\stonybrkc}{Chemistry Department, Stony Brook University, SUNY, Stony Brook, New York 11794-3400, USA}
\newcommand{\stonycrkp}{Department of Physics and Astronomy, Stony Brook University, SUNY, Stony Brook, New York 11794-3400, USA}
\newcommand{\tenn}{University of Tennessee, Knoxville, Tennessee 37996, USA}
\newcommand{\titech}{Department of Physics, Tokyo Institute of Technology, Oh-okayama, Meguro, Tokyo 152-8551, Japan}
\newcommand{\tsukuba}{Institute of Physics, University of Tsukuba, Tsukuba, Ibaraki 305, Japan}
\newcommand{\vandy}{Vanderbilt University, Nashville, Tennessee 37235, USA}
\newcommand{\waseda}{Waseda University, Advanced Research Institute for Science and Engineering, 17 Kikui-cho, Shinjuku-ku, Tokyo 162-0044, Japan}
\newcommand{\weizmann}{Weizmann Institute, Rehovot 76100, Israel}
\newcommand{\wigner}{Institute for Particle and Nuclear Physics, Wigner Research Centre for Physics, Hungarian Academy of Sciences (Wigner RCP, RMKI) H-1525 Budapest 114, POBox 49, Budapest, Hungary}
\newcommand{\yonsei}{Yonsei University, IPAP, Seoul 120-749, Korea}
\affiliation{\abilene}
\affiliation{\banaras}
\affiliation{\barc}
\affiliation{\bnlcoll}
\affiliation{\bnlphys}
\affiliation{\caucr}
\affiliation{\charlesczech}
\affiliation{\chonbuk}
\affiliation{\ciae}
\affiliation{\cns}
\affiliation{\colorado}
\affiliation{\columbia}
\affiliation{\czechtech}
\affiliation{\dapnia}
\affiliation{\elte}
\affiliation{\ewha}
\affiliation{\fit}
\affiliation{\fsu}
\affiliation{\gsu}
\affiliation{\hiroshima}
\affiliation{\ihepprot}
\affiliation{\illuiuc}
\affiliation{\inrras}
\affiliation{\instpasczech}
\affiliation{\isu}
\affiliation{\jinrdubna}
\affiliation{\jyvaskyla}
\affiliation{\kek}
\affiliation{\korea}
\affiliation{\kurchatov}
\affiliation{\kyoto}
\affiliation{\labllr}
\affiliation{\lawllnl}
\affiliation{\losalamos}
\affiliation{\lpc}
\affiliation{\lund}
\affiliation{\maryland}
\affiliation{\mass}
\affiliation{\muenster}
\affiliation{\muhlenberg}
\affiliation{\myongji}
\affiliation{\nagasaki}
\affiliation{\newmex}
\affiliation{\nmsu}
\affiliation{\ornl}
\affiliation{\orsay}
\affiliation{\peking}
\affiliation{\pnpi}
\affiliation{\riken}
\affiliation{\rikjrbrc}
\affiliation{\rikkyo}
\affiliation{\saispbstu}
\affiliation{\saopaulo}
\affiliation{\stonybrkc}
\affiliation{\stonycrkp}
\affiliation{\tenn}
\affiliation{\titech}
\affiliation{\tsukuba}
\affiliation{\vandy}
\affiliation{\waseda}
\affiliation{\weizmann}
\affiliation{\wigner}
\affiliation{\yonsei}
\author{A.~Adare} \affiliation{\colorado}
\author{S.~Afanasiev} \affiliation{\jinrdubna}
\author{C.~Aidala} \affiliation{\mass}
\author{N.N.~Ajitanand} \affiliation{\stonybrkc}
\author{Y.~Akiba} \affiliation{\riken} \affiliation{\rikjrbrc}
\author{H.~Al-Bataineh} \affiliation{\nmsu}
\author{J.~Alexander} \affiliation{\stonybrkc}
\author{A.~Angerami} \affiliation{\columbia}
\author{K.~Aoki} \affiliation{\kyoto} \affiliation{\riken}
\author{N.~Apadula} \affiliation{\stonycrkp}
% ...
\collaboration{PHENIX Collaboration} \noaffiliation

% SPECIAL NOTE:  Please let Brant know, if anyone made significant
%                contributions to the analysis or preparation
%                of this paper, but was not a Member in Good Standing
%                at the time the data was taken.  Such participants
%                may qualify for authorship of this particular paper
%                under clause B or C of the PHENIX Publication Policies:
% http://www.phenix.bnl.gov/phenix/WWW/publish/zajc/sp/publications/publications.htm
%

%-----------------------------------------------------------------------------|

\date{\today}

\begin{abstract}

We present measured $J/\psi$ production rates in $d+$Au collisions at 
$\sqrt{s_{_{NN}}}$ = 200 GeV over a broad range of transverse momentum 
($p_T$ = 0--14 GeV/$c$) and rapidity ($-2.2 < y < 2.2$).  We construct 
the nuclear-modification factor $R_{d{\rm Au}}$ for these kinematics and 
as a function of collision centrality (related to impact parameter for 
the $R_{d{\rm Au}}$ collision).  We find that the modification is largest 
for collisions with small impact parameters, and observe a suppression 
($R_{d{\rm Au}}<1$) for $p_T<4$ GeV/$c$ at positive rapidities. At 
negative rapidity we observe a suppression for $p_T<2$ GeV/$c$ then an 
enhancement ($R_{d{\rm Au}}>1$) for $p_T>2$ GeV/$c$.  The observed 
enhancement at negative rapidity has implications for the observed 
modification in heavy-ion collisions at high $p_T$.

\end{abstract}

	
% For heavy ion papers we usually use just the one above (max is 4)
%%%%%%%%% Examples for p+p and spin papers include:
% PPG031:  \pacs{14.20.Dh, 13.60.Hb, 21.10.Hw, 25.40.Fq}
% PPG050:  \pacs{14.20.Dh, 25.40.Ep, 13.85.Ni, 13.88.+e}
% PPG037:  \pacs{13.85.Qk, 13.20.Fc, 13.20.He, 25.75.Dw}

% It is optional to also add (uncomment):
% \keywords{}

\maketitle

%%========================================================================%%
%%========================================================================%%
%%========================================================================%%
\section{Introduction \label{sec:intro}}

Modifications of quarkonia yields when production takes place in a 
nuclear target, often termed cold-nuclear-matter (CNM) effects, give 
insight into the production and evolution of $q\bar{q}$ pairs. A number 
of effects are predicted to occur in the presence of nuclear matter (for 
a recent review, see~\cite{Brambilla:2010cs}). These include nuclear 
breakup, modification of the parton-distribution functions, initial-state 
parton-energy loss and, more recently, coherent gluon saturation. 
Measuring the production rate of quarkonia in a nuclear environment over 
a broad range of collision energies, and as a function of all kinematic 
variables, is the best way to disentangle these different mechanisms.

The measurement of \jpsi production rates over a broad range of rapidity 
($y$) and transverse momentum (\pt) samples a wide range of parton 
momentum fraction ($x$) and energy transfer ($Q^{2}$), providing a 
simultaneous constraint on the modification of parton-distribution 
functions inside nuclei (nPDF's). The production of \jpsi mesons, which 
at RHIC occurs mainly through gluon fusion, can provide critical input on the 
modification of the gluon distribution, which is probed only indirectly 
by the deep-inelastic scattering (DIS) data that forms the bulk of the 
current constraints on the nPDF parametrizations.

Measuring the \pt distribution of \jpsi production allows access to 
\pt-broadening effects, which are not constrained by 
measurements of the rapidity dependence alone. The 
\pt-broadening effects on quarkonia production at high 
energies are not well constrained by current data. New data for \jpsi 
production over a broad range in \pt is necessary to provide guidance for 
theoretical calculations.

The CNM effects on \jpsi production have been studied in fixed-target 
$p+A$ experiments at SPS, FNAL, and HERA~\cite{Abreu:1998ee, 
Alessandro:2003pi, Alessandro:2003pc, Alessandro:2006jt, Arnaldi:2010ky, 
Leitch:1999ea, Abt:2008ya} spanning the center of mass energy range 
$\sqs\approx 17-42$ GeV. The fixed-target results at midrapidity show 
greater suppression of \jpsi production at lower collision 
energy~\cite{Arnaldi:2010ky}. This has been 
interpreted~\cite{Lourenco:2008sk} as an increase of the nuclear breakup 
of the \jpsi through collisions with nuclei. At lower collision energy 
the crossing time of the nuclei is long enough for the \jpsi to fully 
form. The fully formed \jpsi has an increased probability of interacting 
with other nucleons in the collision, which can cause the breakup of the 
\jpsi into heavy-meson pairs. At higher collision energies it is likely 
that the time required for the \jpsi to fully evolve is as long, or 
longer than the crossing time of the collision. This may result in a 
decrease in the probability of collisions with other nucleons, leading to 
less suppression of the \jpsi production.

The E866~\cite{Leitch:1999ea} and HERA-B~\cite{Abt:2008ya} experiments 
have measured \jpsi production as a function of \pt in fixed target $p+A$ 
experiments. Results are presented in terms of the nuclear-suppression 
factor, $\alpha$, which is obtained assuming that the cross section for 
$p+A$ collisions scales as $\sigma_{pA}=\sigma_{pN}\times A^{\alpha}$, 
where $\sigma_{pN}$ is the proton-nucleon cross section and $A$ is the 
mass number. They find a \pt dependence of $\alpha$, which is similar 
across a range of Feynman-$x$ ($x_F$) and \pt. At $\pt<2$ \gevc they find 
a suppression in the \jpsi production that transitions to an excess in 
the \jpsi production at higher \pt, which is characteristic of multiple 
scattering of the incident parton~\cite{Leitch:1999ea}. It is crucial to 
test these conclusions at the higher energies provided by \dau collisions 
at RHIC in order to better understand the \jpsi production mechanisms.

Measuring, and understanding, the CNM effects on quarkonia production is 
critical to interpreting the results for \jpsi production in 
nucleus-nucleus ($A+A$) collisions. In 1986 Matsui and Satz predicted 
that the suppression of \jpsi production in heavy-ion collisions would be 
a clear signature of the formation of a quark-gluon 
plasma~\cite{Matsui:1986dk}. The Debye color screening of the dense 
medium produced is expected to cause the dissociation of bound states, 
thereby causing a decrease in the observed production. Since then 
suppression of quarkonia production has been observed for a number of 
states, including the \jpsi and $\Upsilon$, over a wide range in 
collision energy~\cite{Arnaldi:2007zz, Adare:2006ns, Chatrchyan:2012np, 
Abelev:2012rv}. However, the interpretation of these results is still 
unclear. Before the modification due to the produced medium can be 
determined, the CNM effects must first be corrected.  This has been 
done at lower energies~\cite{Arnaldi:2007zz}, but accurate data on CNM 
effects are still absent at the higher energies of RHIC and the LHC.

Here we report new high-precision measurements of the \jpsi production as 
a function of \pt and collision centrality in \dau 
collisions at \sqsn = 200 GeV. We also present measurements of the \jpsi 
$R_{d{\rm Au}}$ as a function of \pt, rapidity, and collision 
centrality using data for \jpsi production in \pp collisions published 
in~\cite{Adare:2011vq}. PHENIX has previously measured the \jpsi yield in 
\dau collisions~\cite{Adare:2007gn, Adler:2005ph} with data recorded in 
2003. The data presented here, recorded in 2008, feature an increase in 
statistics of 30--50 times over those used in the previously published 
results, as well as a significant reduction of the systematic 
uncertainties. The rapidity dependence of \jpsi production in \dau 
collisions from this data set has been previously published 
in~\cite{Adare:2010fn}. This paper presents results for the 
\pt dependence of the \jpsi yield from the same data set.

%%========================================================================%%
%%========================================================================%%
%%========================================================================%%
\section{Experimental Apparatus \& Data Sets \label{sec:exp}}

The PHENIX detector~\cite{Adcox:2003zm} comprises three separate 
spectrometers in three pseudorapidity ($\eta$) ranges. Two central arms 
at midrapidity cover \midn and have an azimuthal coverage ($\phi$) of 
$\pi/2$ rad each, while muon arms at backward/forward rapidity cover 
\bacn (Au going direction) and \forn ($d$ going direction), with full 
azimuthal coverage.

In the central arms the \jpsi yield is measured via dielectron decays. 
Charged particle tracks are reconstructed using the drift chamber and pad 
chambers.  Electron candidates are selected by matching charged tracks to 
hits in the ring imaging \v{C}erenkov (RICH) counters and clusters in the 
Electromagnetic Calorimeters (EMCal). In \dau collisions, a charged track 
is identified as an electron by requiring at least two matching RICH 
phototube hits within a radius of $3.4<R[cm]<8.4$ with respect to the 
center defined by the track projection at the RICH. It is also required 
that the position of the EMCal cluster associated to the track projection 
match within $\pm4\sigma$, and that the ratio of the energy deposited in 
the EMCal cluster to the momentum of the tracks matches unity within 
$\pm2.5\sigma$, where $\sigma$ characterizes the momentum dependent width 
of the matching distributions. A further cut of 200 \mevc on the momentum 
of the electron is added to reduce the combinatorial background, since 
the yield of electrons from \jpsi decays observed in data and simulations 
is negligible below 200 \mevc.

At forward and backward rapidity, the \jpsi yield is measured via dimuon 
decays. Muons are identified by matching tracks measured in cathode-strip 
chambers, referred to as the muon tracker (MuTr), to hits in alternating 
planes of Iarocci tubes and steel absorbers, referred to as the muon 
identifier (MuID).  Each muon arm is located behind a thick copper and 
iron absorber that is meant to stop most hadrons produced during the 
collisions, so that the detected muons must penetrate 8 to 11 interaction 
lengths of material in total.

Beam interactions are selected with a minimum-bias (MB) trigger requiring 
at least one hit in each of two beam-beam counters (BBCs) located at 
positive and negative pseudorapidity $3<|\eta|<3.9$. The MB  
selection covers $88 \pm 4\%$ of the total \dau inelastic cross section 
of 2260 mb~\cite{White:2005kp}.

The \dau data sample used in this analysis requires the MB
trigger to be in coincidence with an additional Level-1 trigger. For 
electrons, this is a single electron EMCal RICH trigger (ERT), which 
requires a minimum energy deposited in any 2$\times$2 group of EMCal 
towers, plus an associated hit in the RICH. Two thresholds on the minimum 
EMCal energy, 600 MeV and 800 MeV, were used, each for roughly half of 
the data sample.  For muons, the level 1 trigger requires two tracks 
identified as muon candidates. The trigger logic for a muon candidate 
requires a ``road'' of fired Iarocci tubes in at least 4 planes, 
including the most downstream plane relative to the collision point. 
Additionally, collisions are required to be within $\pm30$cm of the 
center of the interaction region. Collisions in that range see the full 
geometric acceptance of the central arms, and this cut also provides a 
reduction of the systematic uncertainties on the centrality selection 
needed for the data from the muon arms. The data sets sampled via the 
Level-1 triggers represent analyzed integrated luminosities of 62.7 
nb$^{-1}$ (electrons) and 54.0 nb$^{-1}$ (muons) and nucleon-nucleon 
integrated luminosities of 24.7 pb$^{-1}$ and 21.3 pb$^{-1}$ 
respectively.

%====================================================== Table_I
\begin{table}[tbh]
\caption{\label{tab:centrality}
Characterization of the collision centrality for \dau collisions along 
with the correction factor $c$ (see text for details).
}
\begin{ruledtabular}\begin{tabular}{cccc}
Centrality & \mncol & $c$ & $c/\mncol$ \\
\hline
 0--20 \% & 15.1 $\pm$ 1.0 & 0.94  $\pm$  0.01 & 0.062 $\pm$ 0.003 \\
20--40 \% & 10.2 $\pm$ 0.7 & 1.000 $\pm$ 0.006 & 0.098 $\pm$ 0.004 \\
40--60 \% &  6.6 $\pm$ 0.4 & 1.03  $\pm$  0.02 & 0.157 $\pm$ 0.008 \\
60--88 \% &  3.2 $\pm$ 0.2 & 1.03  $\pm$  0.06 & 0.33  $\pm$  0.02 \\
0--100 \% &  7.6 $\pm$ 0.4 & 0.889 $\pm$ 0.001 & 0.117 $\pm$ 0.004 \\
\end{tabular}\end{ruledtabular}
\end{table}

The centrality, which is related to the impact parameter, $b$, of the 
\dau collision is determined using the total charge deposited in the BBC 
located at negative rapidity (Au-going direction). The centrality is 
defined as a percentage of the total charge distribution referenced to 
the greatest charge, $i.e.$ 0--20\% refers to the 20\% of the total charge 
distribution with the greatest charge. On average the 0--20\% centrality 
corresponds to collisions with the smallest $b$.

%%%%%%%%%%%%%%%%%%%%%%%%%%%%%%%%%%%%%%%%%%%%%%%%%%%%%%%% Fig_1
\begin{figure}[thb]
\includegraphics[width=1.0\linewidth]{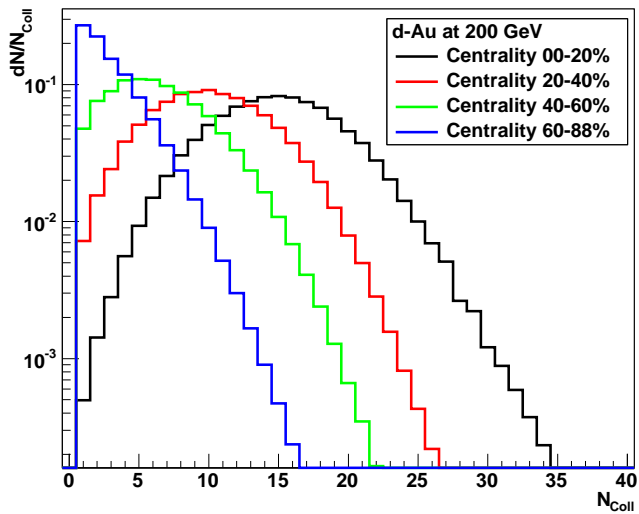}
\caption{(Color online) 
Nucleon-nucleon collision (\Ncoll) distributions for each centrality bin 
obtained using a Glauber model for \dau collisions described in the 
text.}
\label{fig:dAub}
\end{figure}

For each centrality bin the mean number of nucleon-nucleon collisions 
(\mncol) is determined using a Glauber calculation~\cite{Miller:2007ri} 
combined with a simulation of the BBC response (as described 
in~\cite{Adare:2010fn}). The resulting \mncol values for the centrality 
categories used in this analysis are shown in Table~\ref{tab:centrality}. 
The \Ncoll distributions within each centrality bin are shown in 
Fig.~\ref{fig:dAub}. There is a significant overlap between the \Ncoll 
distributions for different centralities.

Also shown in Table~\ref{tab:centrality} is the correction factor $c$, 
which accounts for the correlation between the detection of a \jpsi in 
the final state and an increase in the total charge collected in the 
BBC~\cite{Adare:2007gn}. This correlation affects both the 
MB-trigger efficiency and the determination of the centrality of 
a given collision. The correction factors for each centrality bin are obtained 
within the same Glauber framework as the \mncol values by assuming that 
one of the N binary collisions produces a charge in the BBC that is 
characteristic of a hard-scattering process (the remaining N-1 binary 
collisions maintain a BBC charge distribution characteristic of soft 
scattering processes). The increase in the BBC charge from a hard process 
is tuned using real data.

Since both $c$ and \mncol are calculated in the same Glauber framework 
there are correlations between their uncertainties. These correlations 
are removed in the ratio of $c/\mncol$, which occurs in the calculation 
of \rdau. The resulting values and uncertainties are given in the third 
column of Table~\ref{tab:centrality}. The correction factor for 0--100\% 
centrality contains an additional factor to extrapolate the measured 
yield, which covers only 88\% of all \dau collisions, to 100\% of the 
\dau inelastic cross section, essentially correcting for the efficiency 
of the BBC trigger. This correction is again determined within the 
Glauber framework using the parametrization of the BBC trigger 
efficiency.

%%========================================================================%%
%%========================================================================%%
%%========================================================================%%
\section{\jpsi Analysis and Results in the Midrapidity Region 
\label{sec:anamid}}

The procedure for analyzing the $\jpsi\rightarrow e^+e^-$ signal and the 
results in the central arms are discussed in this section. The extraction 
of the correlated $e^+e^-$ yield is discussed in Sec.~\ref{sec:sigmid}. 
The estimation of the correlated background and losses due to the 
radiative tail in the \jpsi mass distribution is discussed in 
Sec.~\ref{sec:corbgmid}. The estimation of the detector efficiencies is 
described in Sec.~\ref{sec:accmid}. The calculation of the \jpsi 
invariant yield is detailed in Sec.~\ref{sec:invyieldmid}. The \pp 
baseline used in calculating $R_{d{\rm Au}}$ 
is described in Sec.~\ref{sec:ppmid}.

\subsection{Correlated $e^+e^-$ Signal extraction \label{sec:sigmid}}

\begin{figure}[thb]
\includegraphics[width=1.0\linewidth]{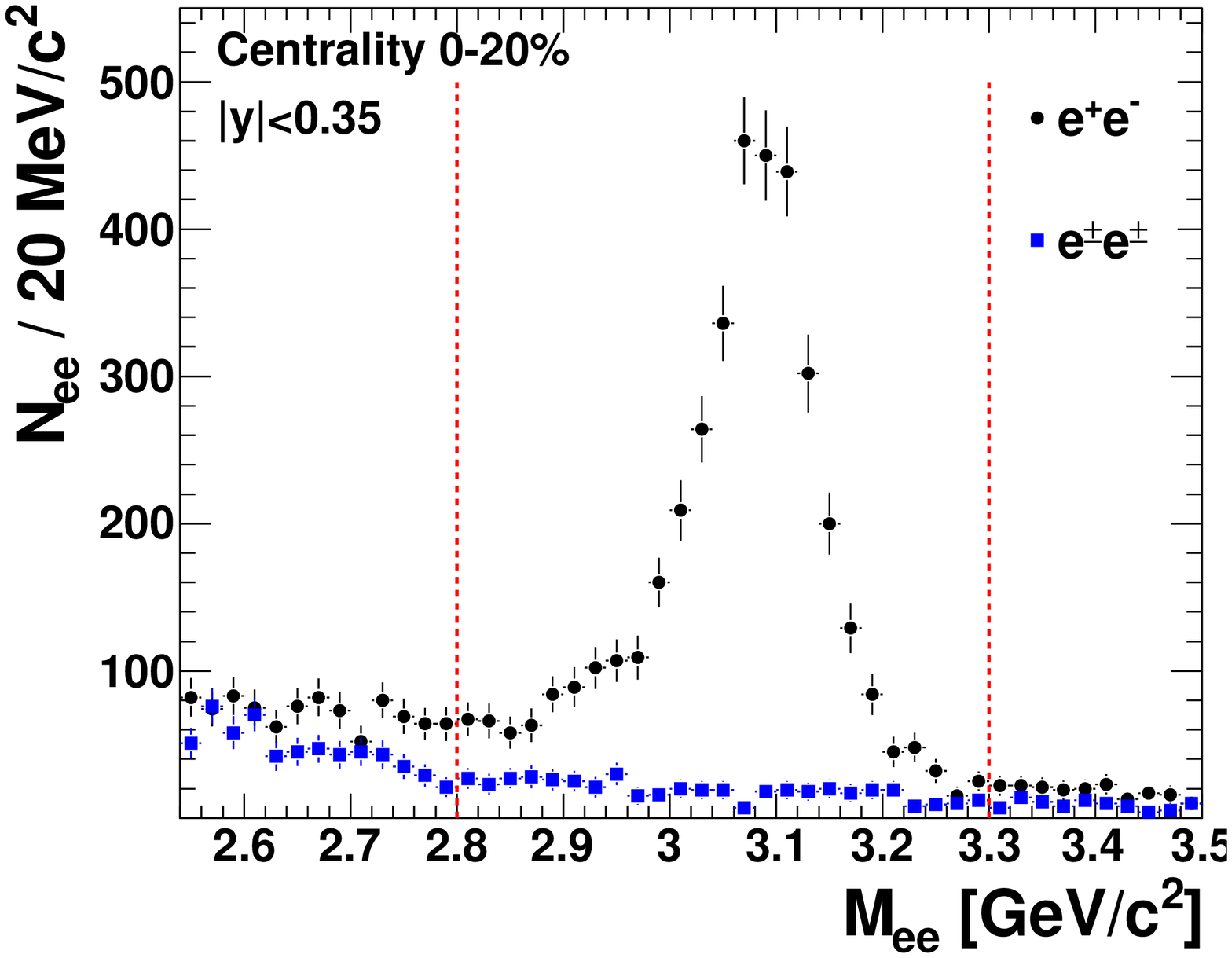}
\includegraphics[width=1.0\linewidth]{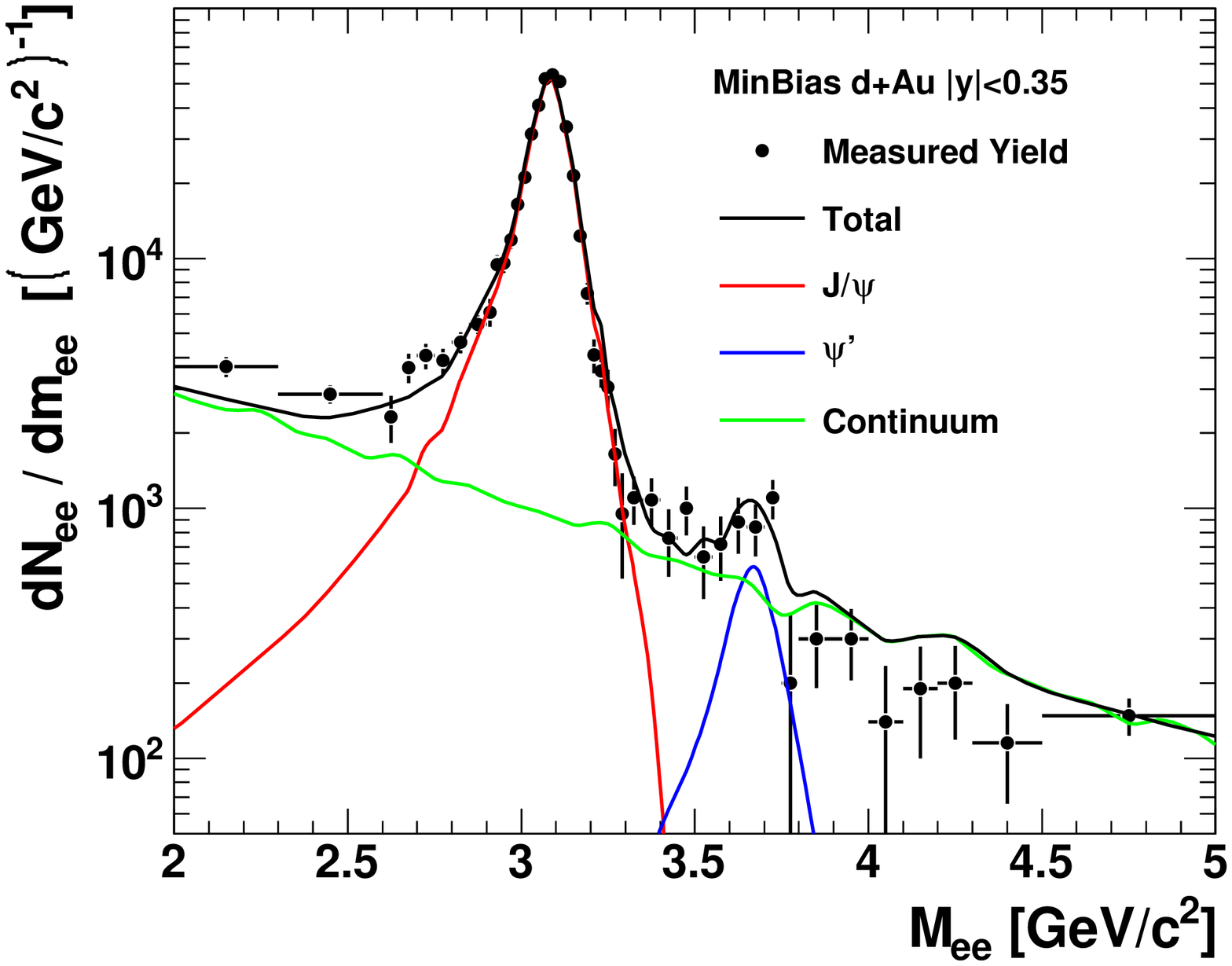}  
\caption{(Color online) 
(top) Invariant mass distribution of unlike-sign (filled circles) and 
like-sign (filled boxes) dielectron pairs in central \dau collisions, 
integrated over \pt and rapidity. Dashed vertical lines represent the 
mass range used to determine the correlated $e^+e^-$ yield. (bottom) 
Correlated dielectron invariant mass distribution for MB \dau 
collisions. The line shapes are those used to extract the continuum and 
radiative tail contributions to the correlated $e^+e^-$ yield in the mass 
range $2.8 < M_{ee}\,[\gevcsq] < 3.3$ .  }
\label{fig:sigmid}
\end{figure}

The $\jpsi\rightarrow e^+e^-$ yield is measured using the invariant mass 
spectrum for all dielectron pairs where at least one of the electrons 
fired the ERT trigger. This selection is necessary to match the 
conditions under which the \jpsi trigger efficiency is calculated (see 
Sec.~\ref{sec:accmid}). An example of the dielectron mass spectrum is 
shown in Fig.~\ref{fig:sigmid} for 0--20\% central collisions. In a 
given bin of \pt, rapidity, and collision centrality, the correlated 
$e^+e^-$ yield ($N_{e^+e^-}$) is determined by counting over a fixed mass 
window of $2.8 < M_{ee}\,[\gevcsq] < 3.3$ the number of unlike-sign 
dielectrons, after the subtraction of the like-sign dielectrons, which 
arise by random association and so are representative of the 
combinatorial background within the unlike sign dielectron distribution. 
This method assumes that the acceptance is the same for $e^-$ and $e^+$, 
which, while untrue at lower masses, is a good assumption in the \jpsi 
mass range. At higher \pt, where statistical precision is limited, the 
yield, along with the statistical uncertainties, are derived from Poisson 
statistics. Assuming both the unlike-sign (foreground) and like-sign 
(background) are independent, and assuming no negative signal, the 
combined distribution \begin{equation} P(s) = 
\sum_{k=0}^{fg}\frac{(bg+fg-k)!}{bg!(fg-k)!}
\frac{1}{2}\left(\frac{1}{2}\right)^{bg+fg-k}\frac{s^ke^{-s}}{k!}, 
\label{eq:sigmid} \end{equation} represents the probability of a signal 
($s$) given a number of unlike-sign dielectrons ($fg$) and a number of 
like-sign dielectrons ($bg$) (see \cite{Adare:2011vq} for derivation). 
The mean and standard deviation of Eq. \ref{eq:sigmid} are then used as 
the yield and uncertainties. A correlated $e^{+}e^{-}$ yield in the mass 
window $2.8 < M_{ee}\,[\gevcsq] < 3.3$ of approximately 8600 is obtained 
across all \pt and collision centralities.

%%%%%%%%%%%%%%%%%%%%%%%%%%%%%%%%%%%%%%%%%%%%%%%%%%%%%%%% Fig_2

\subsection{Estimation of the Correlated Background and Losses Due to the 
Radiative Tail in the \jpsi Mass Distribution 
\label{sec:corbgmid}}

When using the like-sign subtraction method there remains a correlated 
background under the observed \jpsi peak. This background comes mainly 
from open-heavy-flavor decays and Drell-Yan pairs, and must be separated 
from the \jpsi signal of interest. Counting the dielectron signal only 
over a fixed mass window also causes an underestimate of the \jpsi yield 
due to the fraction of the \jpsi line shape that falls outside the mass 
window of choice. These two effects are quantified by using simulated 
particle line shapes fitted to the real data distribution.

The \jpsi and $\psi'$ mesons with uniform distributions in \pt 
($0<\pt\,[\gevc]<12$) and rapidity (\midy) are decayed to $e^+e^-$ and 
the external radiation effects are evaluated using a {\sc geant}-3 based 
model of the PHENIX detector (described in Sec.~\ref{sec:accmid}). While 
a uniform distribution in \pt is unrealistic, the \jpsi rapidity 
distribution is roughly constant within \midy. When used here, the \jpsi 
and \psip line shapes will be compared to \pt integrated data as a 
function of invariant mass only, with a mass resolution fitted to the 
data, and therefore the effect of using a uniform \pt distribution is 
negligible. The line shape for \jpsi radiative decays ($\jpsi\rightarrow 
e^+e^-\gamma$), also called internal radiation, is based on calculations 
of the mass distribution from QED~\cite{Spiridonov:2004mp} convoluted 
with the detector resolution.

Line shapes for the correlated background from heavy-flavor decays along 
with Drell-Yan pairs are simulated using {\sc 
pythia}~\cite{Sjostrand:2006za}. The correlated background from 
heavy-flavor decays comes from semi-leptonic decays of correlated $D\bar{D}$ 
and $B\bar{B}$ (i.e. $D\rightarrow e^{+}+X$ and $\bar{D}\rightarrow 
e^-+X$). The decay electrons from {\sc pythia} are then run through the 
same {\sc geant} simulation of the PHENIX detector to evaluate the 
external radiation effects. These line shapes are generated assuming 
$p+p$ collisions, and no corrections for CNM effects (i.e. application of 
nPDF modifications) are applied to the distributions. We assume that the 
CNM effects on these distributions are likely small and roughly constant 
over the narrow mass window used due to the $x$ values probed. No 
suppression of heavy-flavor production has been observed in \dau 
collisions, and we assume that any suppression, if it exists, does not 
significantly effect the overall line shapes.

The line shapes are then fitted to the \pt and collision centrality 
integrated invariant mass spectrum over the mass range 
$2<M_{ee}\,[\gevcsq]<8$ where the normalizations on the \jpsi, \psip, 
correlated heavy flavor, and DY are free to vary independently. The best 
fit is shown in the quarkonium mass region in Fig.~\ref{fig:sigmid}, 
where the continuum line shape is the combination of correlated $e^+e^-$ 
pairs from $D\bar{D}$, $B\bar{B}$, and DY decays, and the \jpsi and 
$\psi'$ line shapes are the combinations of the line shapes from both 
internal and external radiation effects. Within the mass window 
$2.8<M_{ee}\,[\gevcsq]<3.3$ the correlated continuum contribution 
($\epsilon_{\rm cont}$) is found to be 6.6 $\pm$ 0.2\% and the fraction 
of the \jpsi line shape contained within the mass window ($\epsilon_{\rm 
rad}$) to be 94.3 $\pm$ 0.2\%, where the uncertainties are derived from 
the uncertainty in the fit. The disagreement between the fit and the data 
in the $3.7<M_{ee}\,[\gevcsq]<4.5$ mass range is likely due to the 
inability of the $D\bar{D}$ and $B\bar{B}$ line shapes to match the shape 
of the data at higher mass. However, large changes in the ratio of their 
contributions have only a small effect on the extracted values of 
$\epsilon_{\rm cont}$ and $\epsilon_{\rm rad}$, and this is accounted for 
in the quoted uncertainties.

\subsection{Acceptance and Efficiency Studies \label{sec:accmid}}

The \jpsi acceptance is investigated using a {\sc geant}-3~\cite{GEANT} 
based Monte Carlo model of the PHENIX detector. Dead and malfunctioning 
channels in the detector are removed from both the detector simulation 
and real data. The accuracy of the simulations is tested by comparing 
simulated single electron distributions with those from real data. The 
agreement across the detector and data taking period is determined to be 
within 3.2\%. A conservative estimate, which assumes that the uncertainty 
is correlated for both electrons in a pair, of $2\times3.2\%=6.4\%$ is 
assigned as a systematic uncertainty on the \jpsi acceptance based on the 
quality of the matching between simulations and data.

\begin{figure}[thb]
\includegraphics[width=1.0\linewidth]{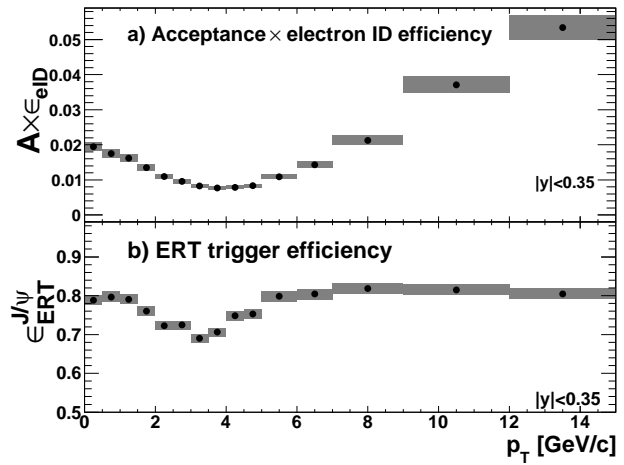}
\caption{
The \jpsi acceptance $\times$ electron ID efficiency (a) and \jpsi ERT 
trigger efficiency (b) as a function of \pt for \midy 
where the shaded boxes represent the systematic uncertainties. }
\label{fig:effmid}
\end{figure}

To determine the \jpsi acceptance, $\jpsi\rightarrow e^+e^-$ decays are 
simulated with uniform distributions in \pt, rapidity ($|y|<0.5$) and 
collision vertex. While distributions uniform in \pt are not realistic, 
the corrections are made over a small \pt bin where the real distribution 
can be approximated as linear. This assumption, and the effect of bin 
sharing, is tested later and taken into account when assigning systematic 
uncertainties. The fraction of \jpsi decays that are reconstructed 
corresponds to the combination of the geometric acceptance and the 
electron ID efficiency ($A\times\epsilon_{\rm eID}$). The resulting 
$A\times\epsilon_{\rm eID}$ is shown as a function of \pt in 
Fig.~\ref{fig:effmid}. It has an average value of 1.5\% in 1 unit of 
rapidity. The dip in $A\times\epsilon_{eID}$ followed by a continual 
increase with \pt marks the transition from the $e^+e^-$ pair at low \pt 
being produced back to back and being detected one in each of the PHENIX 
central arms, to the pair at high \pt being produced in a collinear 
manner and being detected both in the same PHENIX central arm. The low 
point at $\pt\approx 3$ \gevc corresponds to the $e^+e^-$ being produced 
at roughly 90$^{\rm o}$ relative to each other in the lab frame, which 
due to the PHENIX geometry has the lowest probability for detection. The 
electron ID efficiency, which is mainly due to track reconstruction cuts 
used to avoid the misidentification of hadrons as electrons, was cross 
checked using electrons from $\pi^0$ Dalitz decays and $\gamma$ 
conversions as described in~\cite{Adare:2011vq}, and a systematic 
uncertainty of 1.1\% is assigned based on that comparison. The effect of 
momentum smearing on the electrons in simulations, which can cause a 
\jpsi to be reconstructed into a different \pt bin than the one it was 
generated in, was also investigated. The effect was found to be minimal 
for all but the highest \pt bins and an uncertainty of 0.2\% was assigned 
based on a Monte-Carlo study effect and a parametrization of the measured 
momentum resolution for electrons. A combined uncertainty of 6.5\% is 
assigned to the \jpsi $A\times\epsilon_{eID}$ by adding the 
simulation/data matching, eID, and momentum smearing uncertainties in 
quadrature.

The ERT trigger efficiency is evaluated using simulations of \jpsi decays 
and parametrizations of the single electron trigger efficiencies in each 
trigger tile. A MB data sample of single electrons is used to measure the 
\pt dependent efficiency of each 2x2 EMCal trigger tile and each RICH 
trigger tile independently by calculating the fraction of electrons that 
fired the trigger tile compared to all those passing through it. The 
resulting distributions are then fitted with an error(uniform) function 
for each trigger tile in the EMCal(RICH). These functions are then used 
with simulated \jpsi decays to estimate the efficiency of the ERT trigger 
for triggering on $e^+e^-$ pairs from \jpsi decays ($\epsilon_{\rm 
ERT}^{J/\psi}$). The trigger efficiency is evaluated only for simulated 
\jpsi decays for which both electrons passed an acceptance and trigger 
check in order to avoid double counting the acceptance efficiency. This 
procedure is repeated independently for each of the two EMCal trigger 
thresholds used during the run. The \pt dependence of $\epsilon_{\rm 
ERT}^{J/\psi}$ is shown in Fig.~\ref{fig:effmid}, where both ERT 
trigger thresholds have been combined using the relative luminosities of 
each data sample. It has an average value of 77\%. The dip seen at 
$\pt\approx 3$ \gevc is due to the kinematics of the \jpsi decays. In 
that \pt range there is a high probability for the decay electrons to 
have unbalanced momenta, where one of the electrons will have a momentum 
below or near the trigger threshold, resulting in a lower probability for 
triggering on the \jpsi. The effect of the fit function used in the EMCal 
trigger tile efficiencies is investigated by replacing the error function 
with a double-Fermi function. This gives an average change in the \jpsi 
ERT efficiency of 0.31\%. The statistical uncertainty in the trigger tile 
efficiency leads to an uncertainty in the \jpsi ERT efficiency of 1.6\%. 
Summing these uncertainties in quadrature gives a total uncertainty on 
$\epsilon_{\rm ERT}^{J/\psi}$ of 1.6\%, which is heavily dominated by the 
uncertainty in the efficiency of each ERT trigger tile.

The detector occupancy effect is negligible, even in 0--20\% central \dau 
collisions (a finding consistent with previous embedding studies in 
peripheral Cu+Cu~\cite{Adare:2008sh} with similar multiplicities). A 1\% 
systematic uncertainty was assigned based on studies where simulated 
\jpsi decays were embedded into real events. This result agrees well with 
the studies done in \cite{Adare:2007gn}, where a slightly larger 
systematic uncertainty was assigned because of the lower statistical 
precision of the simulations used.

%%%%%%%%%%%%%%%%%%%%%%%%%%%%%%%%%%%%%%%%%%%%%%%%%%%%%%%% Fig_3

\subsection{Invariant Yield Results \label{sec:invyieldmid}}

\begin{figure}[thb]
\includegraphics[width=1.0\linewidth]{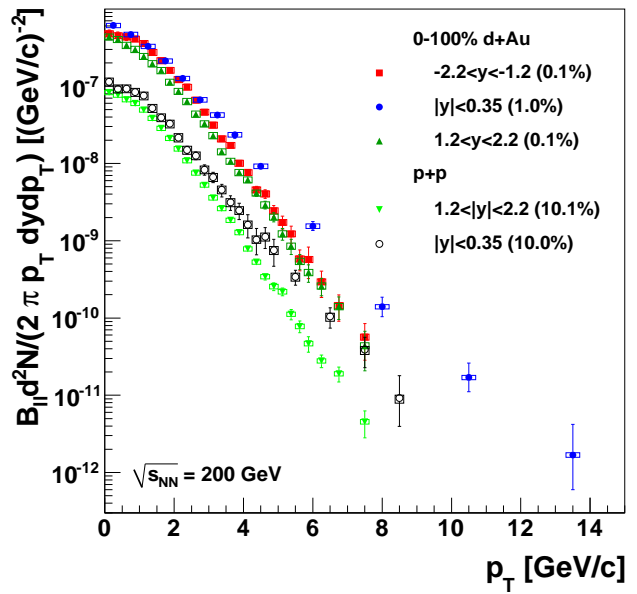}  
\caption{(Color Online)
\jpsi invariant yield as a function of \pt for \pp and 
0--100\% centrality integrated \dau collisions. The type C systematic 
uncertainty for each distribution is given as a percentage in the legend. 
The midrapidity \dau and \pp results are discussed in 
Secs.~\ref{sec:invyieldmid}~\&~\ref{sec:ppmid} while the forward/backward 
rapidity results are discussed in 
Secs.~\ref{sec:invyieldmuon}~\&~\ref{sec:ppmuon}.}
\label{fig:invypt}
\end{figure}

\begin{figure}[thb]
\includegraphics[width=1.0\linewidth]{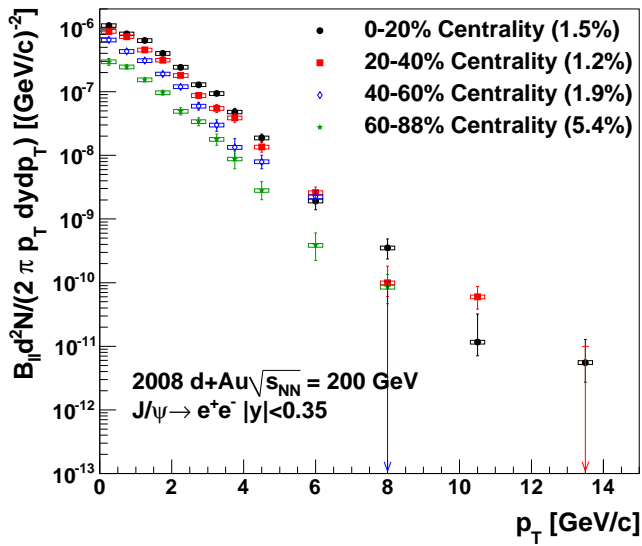}  
\caption{(Color Online)
\jpsi invariant yield as a function of \pt for each 
centrality at \midy. The type C systematic uncertainty for each 
distribution is given as a percentage in the legend.  }
\label{fig:invypt_cent_mid}
\end{figure}

The \jpsi invariant yield in a given rapidity, transverse-momentum, and 
centrality bin is
\begin{equation}
\frac{B_{ll}}{2\pi\pt} \frac{d^{2}N}{dyd\pt} 
= \frac{1}{2\pi\pt\Delta\pt\Delta{y}}
\frac{c N_{J/\psi}}{N_{\rm EVT}\epsilon_{\rm tot}},
\label{eq:invy}
\end{equation}
where $B_{ll}$ is the \jpsi$\rightarrow l^+l^-$ branching ratio, 
$N_{J/\psi}$ is the measured \jpsi yield, $N_{\rm EVT}$ is the number of 
sampled MB events in the given centrality bin, $\Delta{y}$ is the width 
of the rapidity bin, $\Delta{\pt}$ is the width of the \pt
bin, $\epsilon_{\rm tot}=A\times\epsilon_{\rm eID}\ 
\epsilon_{\rm ERT}^{J/\psi}\ \epsilon_{\rm rad}$ and $c$ is the BBC bias 
correction factor described in Sec.~\ref{sec:exp}. At midrapidity 
$N_{J/\psi}=N_{e^+e^-}(1-\epsilon_{\rm cont})$, where $\epsilon_{\rm 
cont}$ is the correlated dielectron continuum contribution in the \jpsi 
mass range.  The 0--100\% centrality integrated \jpsi invariant yield is 
shown as a function of \pt in Fig.~\ref{fig:invypt}, and for four 
centrality bins in Fig.~\ref{fig:invypt_cent_mid}. Here the values 
shown represent the average over the \pt bin and are plotted at the 
center of the bin, as this provides the measured information without 
introducing further systematic uncertainties.

%====================================================== Table_II
\begin{table}[tbh]
\centering
\caption{A summary of the systematic uncertainties at \midy.}
\label{tab:syserrmid}
\begin{ruledtabular}\begin{tabular}{ccc}
Source & Value & Type \\
\hline
Embedding & 1.0\% & C \\
$\epsilon_{\rm rad}$ & 0.2\% & C \\
$\epsilon_{\rm cont}$ & 0.2\% & C \\
$c$ (Invariant yield only) &0.1--5.8\% & C\\
$c/\mncol$ (\rdau only) & 3--6\% & C \\ 
$A\times\epsilon_{\rm eID}$. & 6.5\% & B \\
$\epsilon_{\rm ERT}^{J/\psi}$. & 1.6\% & B \\
Stat. Uncertainty on & & \\
the correlated $e^+e^-$ yield & & A \\
\end{tabular}\end{ruledtabular}
\end{table}

A summary of all the relevant systematic uncertainties at midrapidity is 
shown in Table~\ref{tab:syserrmid}, along with their classification into 
Type A, B, or C uncertainties. Type A represents uncertainties that are 
uncorrelated from point to point, Type B represents uncertainties that 
are correlated from point to point, and Type C represents uncertainties 
in the overall normalization.

%%%%%%%%%%%%%%%%%%%%%%%%%%%%%%%%%%%%%%%%%%%%%%%%%%%%%%%% Fig_4

%%%%%%%%%%%%%%%%%%%%%%%%%%%%%%%%%%%%%%%%%%%%%%%%%%%%%%%% Fig_5

\subsection{\pp Baseline \label{sec:ppmid}}

The \pp baseline used to calculate $R_{d{\rm Au}}$ is 
extracted from 2006 data published in~\cite{Adare:2011vq}.  The 
integrated luminosity was 6.2$\pm$0.6 pb$^{-1}$. In the analysis, 
described in detail in~\cite{Adare:2011vq}, the effect of the \jpsi 
polarization on the \jpsi acceptance is included. This effect is not 
included in the \dau result presented here due to a lack of knowledge of 
the effects of a nuclear target on the \jpsi polarization. The \jpsi 
polarization is therefore assumed to be zero. To remain consistent, this 
effect is removed from the \pp baseline as well, so that, assuming the 
polarization does not change drastically between \pp and \dau, the 
effects will cancel in the nuclear ratio, \rdau. The \pp invariant yields 
as a function of \pt used in this work, shown in Fig.~\ref{fig:invypt}, 
have been converted from the invariant cross sections published 
in~\cite{Adare:2011vq} using an inelastic cross section of 42 mb.

%%========================================================================%%
%%========================================================================%%
%%========================================================================%%
\section{\jpsi Analysis and Results in the Forward/Backward Rapidity Region}
\label{sec:anabacfor}

The procedure for analyzing the $\jpsi\rightarrow\mu^+\mu^-$ signal at 
backward and forward rapidity in the muon arms is discussed in this 
section. The procedures are similar to those detailed 
in~\cite{Adare:2011vq}, with only a brief summary presented here, except 
where there are differences. As in~\cite{Adare:2011vq}, the rapidity 
region of the forward muon arm used in the analysis was truncated to 
\fory to match the rapidity coverage of the backward muon arm. The 
extraction of the raw \jpsi yield is discussed in Sec.~\ref{sec:sigmuon}. 
The estimation of the detector efficiencies is described in 
Sec.~\ref{sec:accmuon}. The calculation of the \jpsi invariant yield is 
detailed in Sec.~\ref{sec:invyieldmuon}. The \pp baseline used in 
calculating $R_{d{\rm Au}}$ is described in~Sec.\ref{sec:ppmuon}.

\subsection{$\jpsi\rightarrow\mu^+\mu^-$ Signal extraction}
\label{sec:sigmuon}

At forward and backward rapidity, the invariant mass distribution is 
calculated for all unlike-sign dimuons in events that pass the trigger 
requirements described in Sec.~\ref{sec:exp}. The combinatorial 
background is estimated from the invariant mass distribution formed by 
pairing unlike-sign muon candidates from different events. This is done 
to reduce the background statistical uncertainty below what is possible 
by subtracting like sign pairs from the same event, and is needed because 
the signal to background present at forward/backward rapidity is smaller 
than at midrapidity. The mixed event muon pairs are required to have 
vertices that differ by no more than 3 cm in the beam direction. The 
mixed event spectrum is normalized by the factor
\begin{eqnarray}
\alpha=\frac{\sqrt{(N_{\mu^+\mu^+}^{\rm same})
(N_{\mu^-\mu^-}^{\rm same})}}{\sqrt{(N_{\mu^+\mu^+}^{mixed})
(N_{\mu^-\mu^-}^{mixed})}},
\end{eqnarray}
where $N_{\mu\mu}^{\rm same}$ and $N_{\mu\mu}^{mixed}$ are the number of 
pairs formed from two muons in the same or in mixed events, respectively.

The remaining correlated dimuon mass distribution after the subtraction 
of the mixed event combinatorial background contains dimuons from \jpsi 
and \psip decays, as well as correlated dimuons from heavy-flavor decays 
and Drell-Yan pairs. Due to the momentum resolution of the detector, 
there is no clean discrimination between the \jpsi and \psip in the mass 
distribution. However the \psip contribution is expected to be negligible 
in the mass window of interest.

\begin{figure}[thb]
\includegraphics[width=1.0\linewidth]{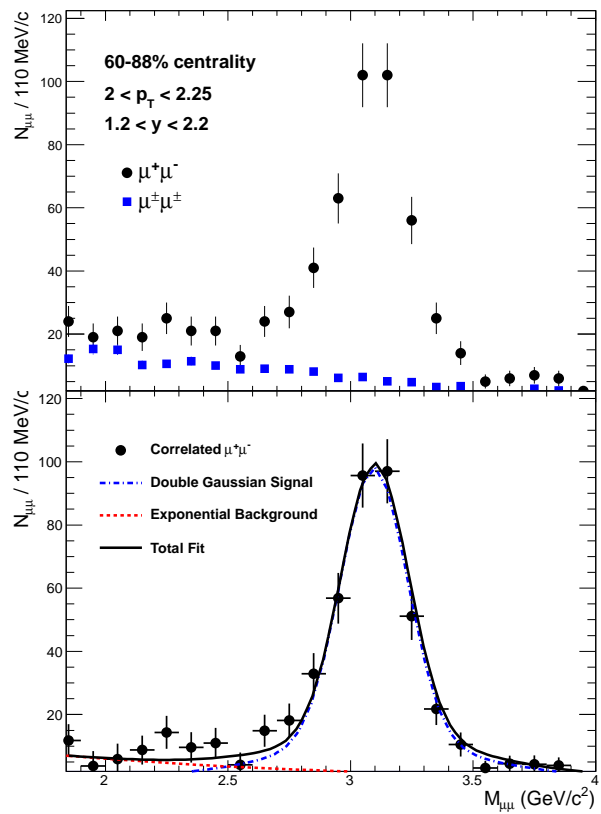} 
\caption{(Color Online) 
(top) Invariant mass distribution of unlike-sign (filled circles) and 
like-sign (filled boxes) dimuon pairs for $2<\pt<2.25$ \gevc at forward 
rapidity and 60--88\% central events. (bottom) Invariant mass distribution 
of correlated dimuon pairs after the subtraction of the combinatorial 
background. The solid line represents the fit to the invariant mass 
distribution, which includes the double Gaussian signal component 
(dot-dashed line) and exponential background (dotted line). }
\label{fig:sigfor}
\end{figure}

A function consisting of an exponential component combined with two 
Gaussian distributions, which are used to better reproduce the mass 
resolution present in the muon arms, was used to fit the dimuon mass 
distribution, convolved with a function to account for the variation in 
acceptance over the invariant mass range. An example of the fitted mass 
distribution is shown in Fig.~\ref{fig:sigfor}. Both the \jpsi 
component of the fit, and direct counting after the subtraction of the 
fitted exponential background, are used to evaluate the \jpsi yield. The 
difference between the two methods is taken as a Type A systematic 
uncertainty. This uncertainty is typically small ($\approx2$\%) but can 
be significantly larger at high \pt where there are fewer counts. 
Measured \jpsi yields of approximately 38000 and 42000 are obtained at 
backward and forward rapidity, respectively.

%%%%%%%%%%%%%%%%%%%%%%%%%%%%%%%%%%%%%%%%%%%%%%%%%%%%%%%% Fig_6

\subsection{Acceptance and Efficiency Studies \label{sec:accmuon}}

Studies of the response of the muon arm spectrometers to dimuons from 
\jpsi decays are performed using a tuned {\sc geant}3-based simulation of 
the muon arms, coupled with a MuID trigger emulator. The MuID 
panel-by-panel efficiencies are estimated using the fraction of 
reconstructed roads in real data. Where statistics are limited, the 
operational history of each channel recorded during the run was used to 
estimate the efficiency. A systematic uncertainty of 4\% is assigned to 
the MUID efficiency based on this comparison. Charge distributions in 
each part of the MuTr observed in real data, along with dead channels and 
their variation with time over the run, are used to give an accurate 
description of the MuTr efficiency within the detector simulation.

The \jpsi acceptance $\times$ efficiency ($A\times\epsilon$) evaluation 
uses a {\sc pythia} simulation with several parton distributions as input 
to account for the unknown underlying rapidity dependence of the \jpsi 
yield.  A 4\% systematic uncertainty is assigned based on changes in the 
input parton distributions. A systematic uncertainty of 6.4(7)\% on the 
\jpsi yield is assigned to the backward(forward) rapidity due to the 
uncertainties in the acceptance x efficiency determination method itself.

\subsection{Invariant Yield Results \label{sec:invyieldmuon}}

%====================================================== Table_III
\begin{table}[tbh]
\caption{\label{tab:syserrmuon}
   The dominant systematic errors at \muony.
}
\begin{ruledtabular}\begin{tabular}{ccc}

Source & Value (S/N) & Type \\
\hline
$c$ (Invariant yield only) & 0.1-5.8\% & C\\
$c/\mncol$ (\rdau only) & 3-6\% & C \\ 
MC Input Distributions. & 4\% & B \\
MuTr Efficiency & 2\% & B \\
MUID Efficiency & 4\% & B \\
Acceptance & 6.4/7\% & B \\
Fit Type & $\approx2$\% & A \\
Stat. Uncertainty on & & \\
the measured \jpsi yield & & A \\

\end{tabular}\end{ruledtabular}
\end{table}

\begin{figure}[thb]
\includegraphics[width=1.0\linewidth]{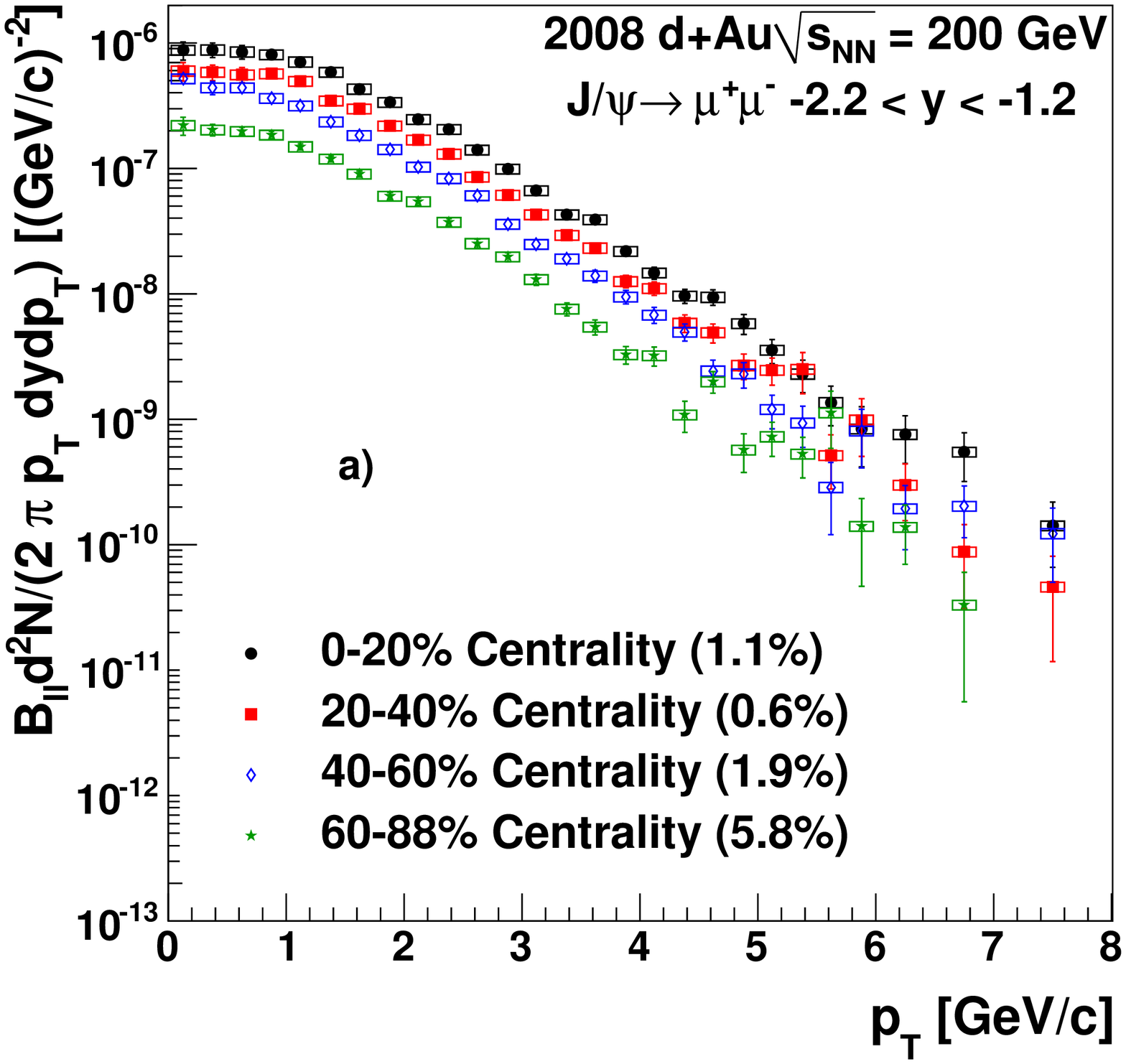}  
\includegraphics[width=1.0\linewidth]{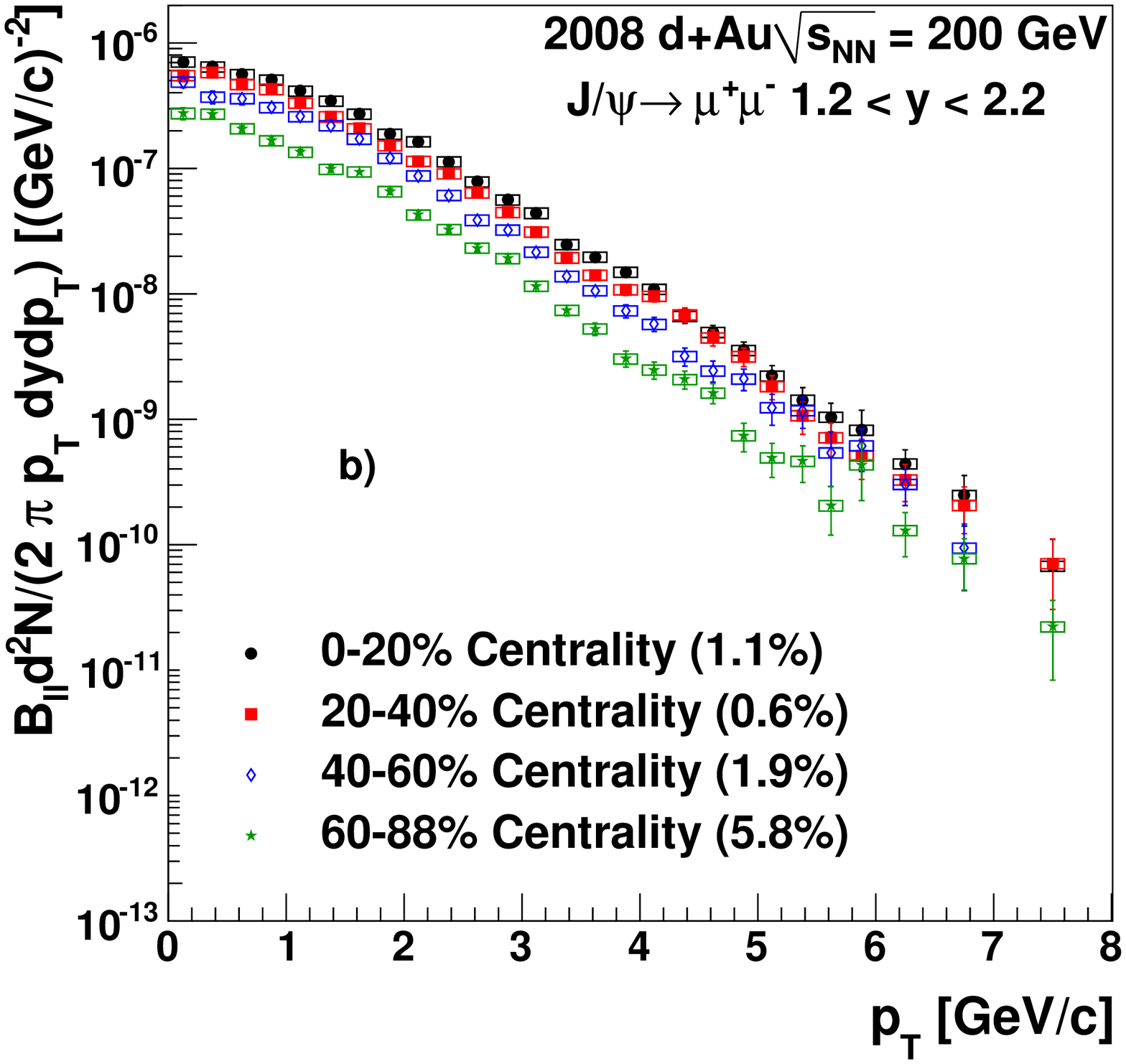}  
\caption{(Color Online)
\jpsi invariant yield as a function of \pt for each 
centrality for a) \bacy and b) \fory. The type C systematic uncertainty 
for each distribution is given as a percentage in the legend. }
\label{fig:invypt_cent_bacfor}
\end{figure}

The \jpsi invariant yield at backward/forward rapidity is calculated 
using Eq.~\ref{eq:invy}, where $\epsilon_{\rm tot}=A\times\epsilon$. A 
summary of the systematic uncertainties is given in 
Table~\ref{tab:syserrmuon}. The backward and forward 0--100\% 
centrality-integrated \jpsi invariant yields are shown as a function of 
\pt in Fig.~\ref{fig:invypt}, while the \jpsi invariant yields are 
shown as a function of \pt in each centrality bin in 
Fig.~\ref{fig:invypt_cent_bacfor}.

%%%%%%%%%%%%%%%%%%%%%%%%%%%%%%%%%%%%%%%%%%%%%%%%%%%%%%%% Fig_7

\subsection{\pp Baseline \label{sec:ppmuon}}

The \pp baseline used to calculate $R_{d{\rm Au}}$ is 
extracted from a combined analysis of data taken in 2006 and 2008, 
published in~\cite{Adare:2011vq}. The combined integrated luminosity was 
9.3$\pm$0.9 pb$^{-1}$. As discussed in Sec.~\ref{sec:ppmid}, the effect 
of the \jpsi polarization on the \jpsi acceptance is removed from the 
results used here. The \jpsi invariant yield in \pp collisions at 
forward/backward rapidity used here is shown as a function of \pt in 
Fig.~\ref{fig:invypt} for convenience, where we have converted from the 
invariant cross sections published in~\cite{Adare:2011vq} using an 
inelastic cross section of 42 mb.

%%========================================================================%%
%%========================================================================%%
%%========================================================================%%
\section{Calculation of \mptsq \label{sec:mptsq}}

The \mptsq is calculated for each of the \jpsi invariant yields presented 
in Secs.~\ref{sec:invyieldmid} and~\ref{sec:invyieldmuon}, and the 
resulting values are shown in Table~\ref{tab:meanpt}.

%====================================================== Table_IV
\begin{table}[tbh]
\caption{
\mptsq results for \pp and \dau collisions where the first quoted 
uncertainty corresponds to the type A uncertainties and the second 
corresponds to the type B uncertainties.}
\label{tab:meanpt} 
\begin{ruledtabular}\begin{tabular}{cccc}
System & y range & Centrality & \mptsq [GeV$^2$/$c^2$] \\
\hline
\pp  & \muony &         & 3.64$\pm$0.03$\pm$0.06 \\
\pp  & \midy  &         & 4.46$\pm$0.14$\pm$0.18 \\
\\
\dau & \bacy  & 0--100\% & 4.09$\pm$0.06$\pm$0.09 \\
\dau & \midy  & 0--100\% & 5.10$^{+0.12}_{-0.10}\pm$0.11 \\
\dau & \fory  & 0--100\% & 4.05$\pm$0.05$\pm$0.10 \\
\\
\dau & \bacy &  0--20\% & 4.22$\pm$0.08$\pm$0.09 \\
\dau & \bacy & 20--40\% & 4.06$\pm$0.08$\pm$0.09 \\
\dau & \bacy & 40--60\% & 4.01$\pm$0.09$\pm$0.09 \\
\dau & \bacy & 60--88\% & 3.92$\pm$0.10$\pm$0.09 \\
\\
\dau & \midy &  0--20\% & 5.24$^{+0.19}_{-0.16}$$\pm$0.10 \\
\dau & \midy & 20--40\% & 5.27$^{+0.22}_{-0.19}$$\pm$0.12 \\
\dau & \midy & 40--60\% & 5.08$^{+0.29}_{-0.26}$$\pm$0.16 \\
\dau & \midy & 60--88\% & 4.60$^{+0.30}_{-0.24}$$\pm$0.15 \\
\\
\dau & \fory &  0--20\% & 4.15$\pm$0.06$\pm$0.10 \\
\dau & \fory & 20--40\% & 4.13$\pm$0.07$\pm$0.11 \\
\dau & \fory & 40--60\% & 3.94$\pm$0.07$\pm$0.10 \\
\dau & \fory & 60--88\% & 3.80$\pm$0.08$\pm$0.10 \\
\end{tabular}\end{ruledtabular}
\end{table}

Unlike in previous analyses~\cite{Adare:2007gn}, where the \mptsq was 
calculated for $\pt\leq 5$ \gevc due to statistical limitations at high 
\pt, here we have calculated the \mptsq over the full \pt range. First 
the \mptsq was calculated numerically up to the \pt limits of the 
measured distribution $\left(\mptsqmax\right)$. The correlated 
uncertainty was propagated to \mptsqmax by sampling the type B 
uncertainty distributions of the first and last \pt point of the 
invariant yield, and assuming a linear correlation in between. For a more 
detailed description of this procedure see Appendix~\ref{sec:typeBmptsq}.

To account for the differences in the \pt limits of the various 
distributions, the \mptsqmax value was corrected to the \pt range from 
zero to infinity. This was done by fitting the distribution with a 
modified Kaplan function of the form
\begin{equation}
f(\pt)=p_0\left(1-\left(\frac{\pt}{p_1}\right)^2\right)^{p_2}
\label{eq:modkaplan1}
\end{equation}
where each parameter was free to vary. The ratio
\begin{equation} 
k=\frac{\mptsq[0,\infty]}{\mptsq[0,\pt^{\rm max}]}
\label{eq:mptsq_corr1}
\end{equation}
was then calculated from the fit and applied to the numerically 
calculated \mptsqmax. In all cases the correction factor was small 
($k<1.03$), and an uncertainty in the correction factor based on the fit 
uncertainty is included in the Type B uncertainties shown in 
Table~\ref{tab:meanpt}. For a more detailed description of this 
procedure, including the fit results and the calculated values of $k$ see 
Appendix~\ref{sec:detailmptsq}.

The \mptsq for \pp collisions was previously published in 
\cite{Adare:2011vq}. But we report the result here with the effect of the 
\jpsi polarization on the acceptance removed. The results are in good 
agreement with those presented in \cite{Adare:2011vq}, and are shown in 
Table~\ref{tab:meanpt}.

\begin{figure}[thb]
\includegraphics[width=1.0\linewidth]{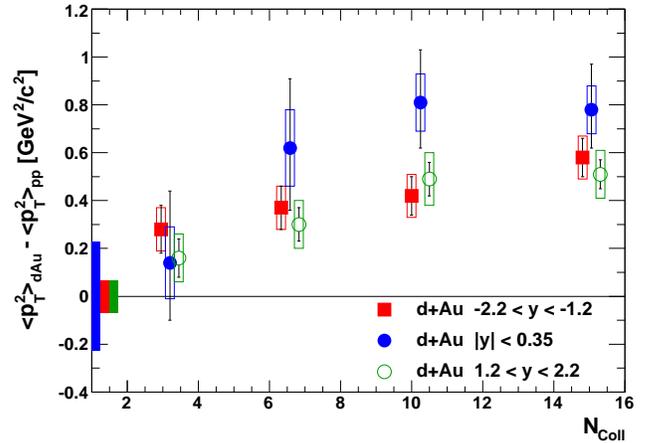}
\caption{(Color Online)
The difference between the \jpsi \mptsq in \dau and \pp collisions as a 
function of \Ncoll in \dau collisions. The boxes drawn at 
$\Delta\mptsq=0$ represent the combined statistical and systematic 
uncertainties from the \pp calculation.}
\label{fig:mptsqdau-pp}
\end{figure}

Figure~\ref{fig:mptsqdau-pp} shows 
$\Delta\mptsq=\mptsq_{dAu}-\mptsq_{pp}$ as a function of \Ncoll. There is 
a broadening in the \pt distribution with respect to \pp, which increases 
with \Ncoll, and is similar at forward and backward rapidities. We 
observe a larger increase in the \pt broadening at midrapidity. However, 
this observation is tempered by the relatively large uncertainties 
present in the data.

%%%%%%%%%%%%%%%%%%%%%%%%%%%%%%%%%%%%%%%%%%%%%%%%%%%%%%%% Fig_8

%%========================================================================%%
%%========================================================================%%
%%========================================================================%%
\section{The \jpsi $R_{d{\rm Au}}$ \label{sec:nucmod}}

%%%%%%%%%%%%%%%%%%%%%%%%%%%%%%%%%%%%%%%%%%%%%%%%%%%%%%%% Fig_9
\begin{figure}[thb]
\includegraphics[width=1.0\linewidth]{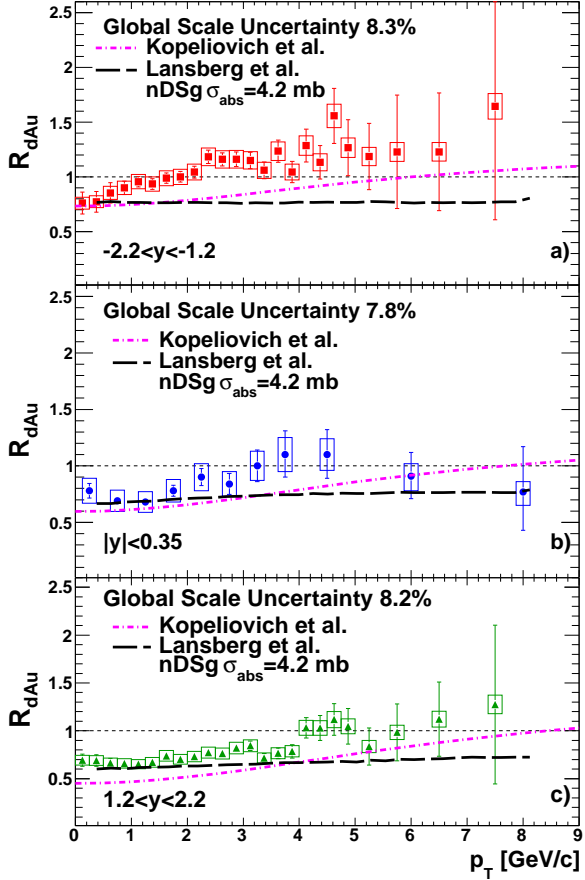}
\caption{(Color Online)
\jpsi nuclear modification factor, \rdau, as a function of \pt
for (a) backward rapidity, (b) midrapidity, and (c) forward 
rapidity 0--100\% centrality integrated \dau collisions. Curves are model 
calculations detailed in Sec.~\ref{sec:rdaumodel}.}
\label{fig:rdauptmb}
\end{figure}

To quantify the \dau cold nuclear matter effects, the \jpsi \rdau 
is calculated for a given \pt, $y$, and centrality bin as:
\begin{equation}
\rdau(i) = \frac{c}{\ncol{i}}\frac{d^2N_{J/\psi}^{d+\rm Au}(i)/dyd\pt}
{d^2N_{J/\psi}^{p+p}/dyd\pt},
\label{eq:rdau}
\end{equation}
where $d^2N_{J/\psi}^{d+\rm Au}(i)/dyd\pt$ is the \dau invariant yield 
for the $i^{\rm th}$ centrality bin, $d^2N_{J/\psi}^{p+p}/dyd\pt$ is the 
\pp invariant yield for the same \pt and $y$ bin, 
and \ncol{i} is the average number of binary collisions for the given 
centrality bin, as listed in Table \ref{tab:centrality}.

\begin{figure}[thb]
\includegraphics[width=1.0\linewidth]{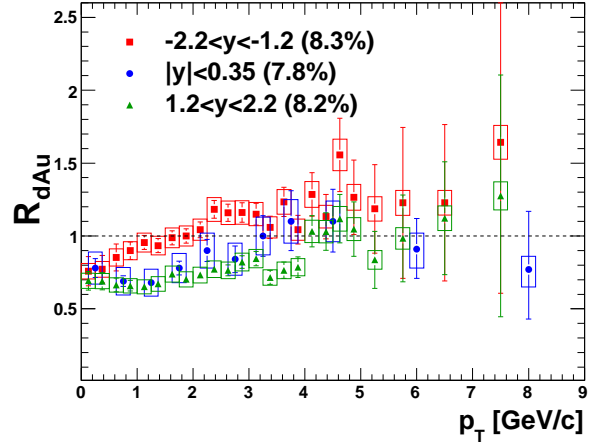}
\caption{(Color Online)
\jpsi \rdau, as a function of \pt 
for 0--100\% centrality integrated \dau collisions at each 
rapidity. The Type C systematic uncertainty for each distribution is 
given as a percentage in the legend.}
\label{fig:mbrdauptall}
\end{figure}

The 0--100\% centrality integrated \jpsi \rdau as a function of 
\pt is shown in Fig.~\ref{fig:rdauptmb} for each of the three 
rapidity regions.  The numerical values can be found in 
Table~\ref{tab:mbrdaubac},~\ref{tab:mbrdaumid}, and~\ref{tab:mbrdaufor} 
for backward, mid and forward rapidity, respectively. 
Figure~\ref{fig:rdauptmb} shows a different behavior for \rdau at 
backward (\bacy) as opposed to mid (\midy) and forward (\fory) 
rapidities. At backward rapidity, the \rdau is suppressed only at the 
lowest \pt, with a rapid increase to $\rdau=1.0$ at $\pt\approx1.5$ 
\gevc. The mid and forward rapidity data, on the other hand, exhibit a 
similar level of suppression at the lowest \pt, but a much more gradual 
increase in \rdau with \pt, increasing to $\rdau=1.0$ only at 
$\pt\approx4.0$ \gevc.  Figure~\ref{fig:mbrdauptall} shows the same 
0--100\% \rdau vs \pt at all rapidities overlaid. It is striking that the 
shape and absolute scale for the mid and forward rapidity data is nearly 
consistent across the entire \pt range of the data.

%%%%%%%%%%%%%%%%%%%%%%%%%%%%%%%%%%%%%%%%%%%%%%%%%%%%%%%% Fig_10

Due to the statistical limitations of the data at high \pt, it is unclear 
from Fig.~\ref{fig:rdauptmb} whether the \rdau increases significantly 
above one. To investigate the high-\pt behavior of the \rdau at each 
rapidity, the average \rdau was calculated for $\pt>4$ \gevc by fitting 
each distribution with a constant. The results are shown in 
Table~\ref{tab:highptrdau} along with the fit uncertainties, which take 
into account only the type A uncertainties on the data. Since the type B 
uncertainties are roughly consistent in the fit range, we have chosen 
here to add the average type B uncertainty for $\pt>4$ \gevc in 
quadrature with the type C uncertainty. We find that at mid and forward 
rapidity the average \rdau for $\pt>4$ \gevc is consistent with 1.0, 
while at backward rapidity the average \rdau is greater than 1.0.

The production of a \jpsi at forward rapidity in $A+A$ collisions 
involves a low-$x$ gluon colliding with a high-$x$ gluon. The symmetry 
due to identical colliding nuclei results, essentially, in the folding of 
the forward and backward rapidity \rdau. The production of a \jpsi at 
midrapidity results, essentially, in the folding of the midrapidity \rdau 
with itself. This picture is simplistic and leaves out many details, but 
it gives some expectation for the result of the modification of \jpsi 
production in $A+A$ collisions due to CNM effects. If we extrapolate the 
observed behavior of \rdau to the modification of \jpsi's produced at 
forward rapidity in $A+A$ collisions, we would expect a \raa contribution 
from CNM effects to be similar to, or greater than, 1.0 at high \pt and a 
modification similar to 1.0 at midrapidity. The observation at 
midrapidity of a \jpsi \raa in Cu+Cu collisions that exceeds, but is 
consistent with, 1.0 at high \pt~\cite{Abelev:2009qaa} may therefore be 
largely accounted for by the contribution from CNM effects. Further work 
is needed to understand the detailed propagation of measured results in 
\dau collisions to an expected CNM contribution in $A+A$ collisions 
before this can be fully understood.

%====================================================== Table_V
\begin{table}[tbh]
\caption{
The average 0--100\% \rdau for $\pt>4$ \gevc where the first quoted 
uncertainty corresponds to the fit uncertainty and the second corresponds 
to the combined type B and C systematic uncertainties.}
\label{tab:highptrdau} 
\begin{ruledtabular}\begin{tabular}{cc}

Rapidity & \rdau($\pt>4$ \gevc) \\
\hline
\bacy & 1.27$\pm$0.06$\pm$0.11 \\
\midy & 0.97$\pm$0.14$\pm$0.16 \\
\fory & 1.03$\pm$0.06$\pm$0.11 \\

\end{tabular}\end{ruledtabular}
\end{table}

\begin{figure}[thb]
\includegraphics[width=1.0\linewidth]{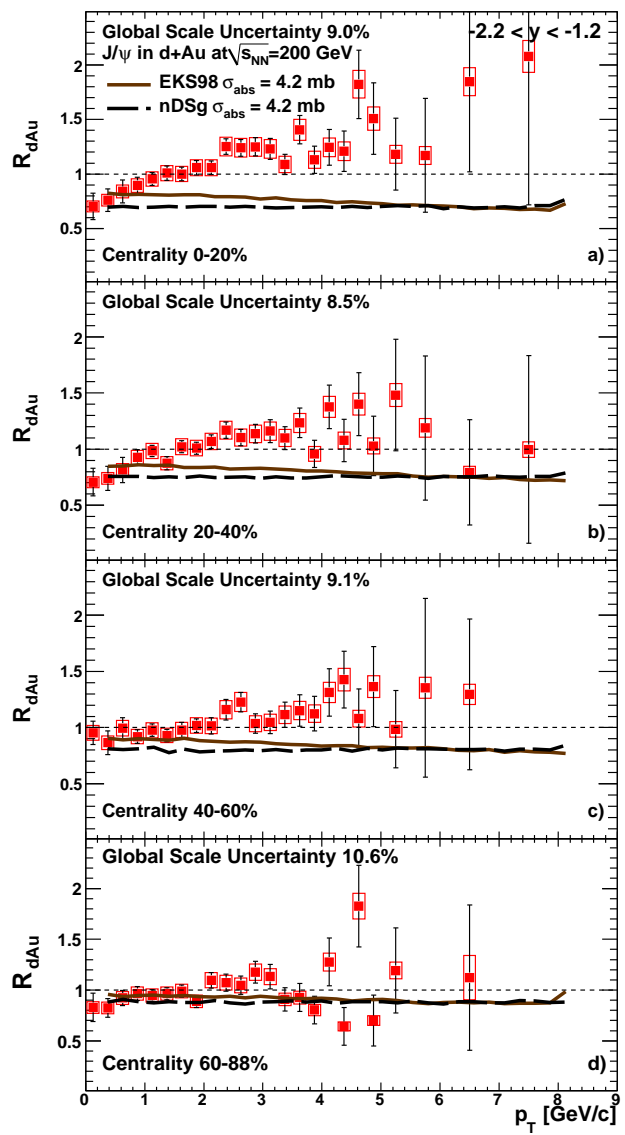}
\caption{(Color Online)
$J/\psi\rightarrow \mu^+\mu^-$ \rdau, as a 
function of \pt for a) central, b) midcentral, c) 
midperipheral, and d) peripheral events at \bacy. The 60--88\% \rdau 
point at $\pt=5.75$ \gevc has been left off the plot, as it is above the 
plotted range and has very large uncertainties, however it is included in 
Table~\ref{tab:rdaucent3}. Curves are calculations by Lansberg et 
al.~\cite{Ferreiro:2012zb} discussed in the text. }
\label{fig:rdauptcent_bac}
\end{figure}

\begin{figure}[thb]
\includegraphics[width=1.0\linewidth]{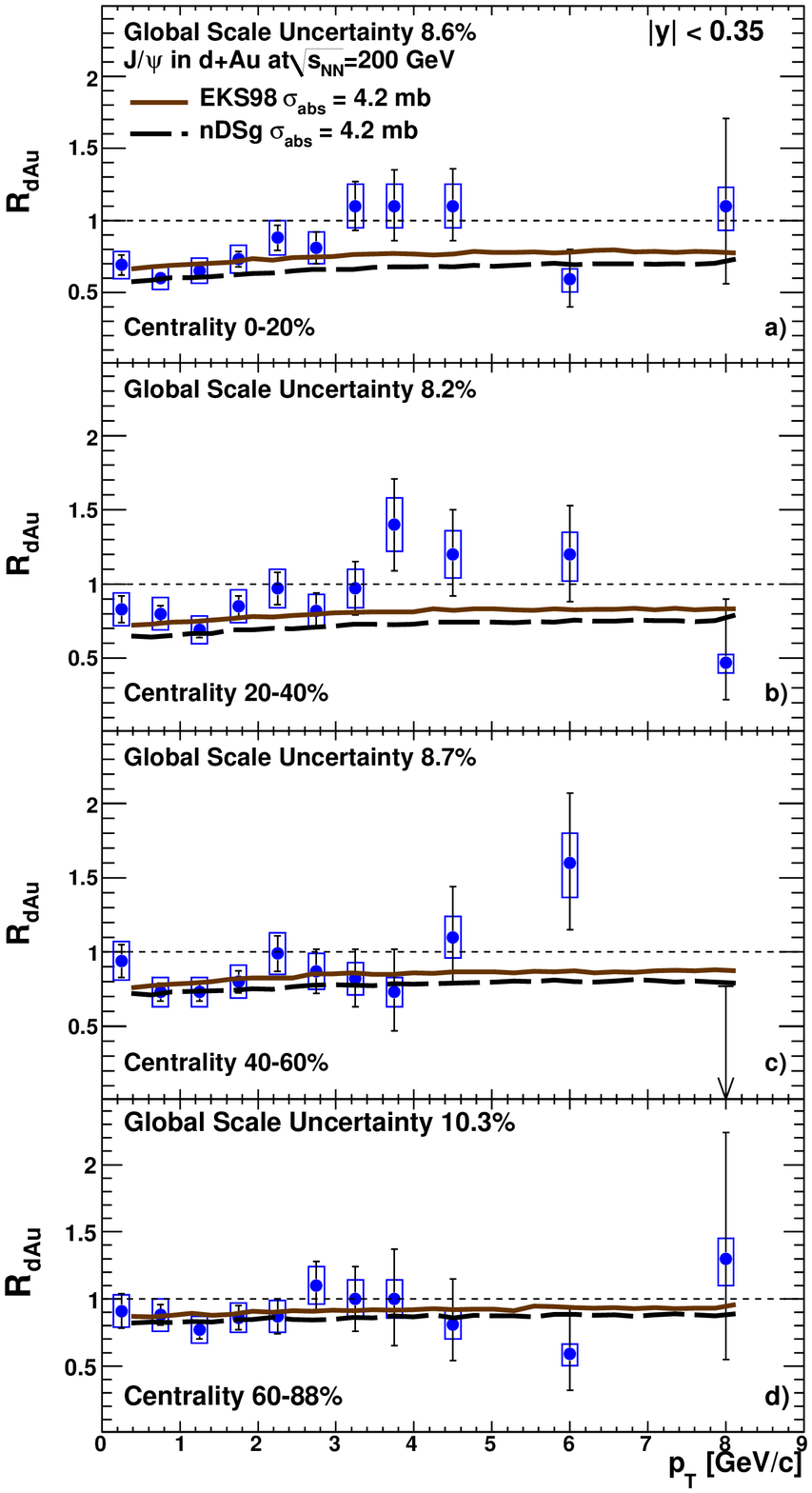}
\caption{(Color Online)
$J/\psi\rightarrow e^+e^-$ \rdau, as a 
function of \pt for a) central, b) midcentral, c) 
midperipheral, and d) peripheral events at \midy. Curves are calculations 
by Lansberg et al.~\cite{Ferreiro:2012zb} discussed in the text.  }
\label{fig:rdauptcent_mid}
\end{figure}

\begin{figure}[thb]
\includegraphics[width=1.0\linewidth]{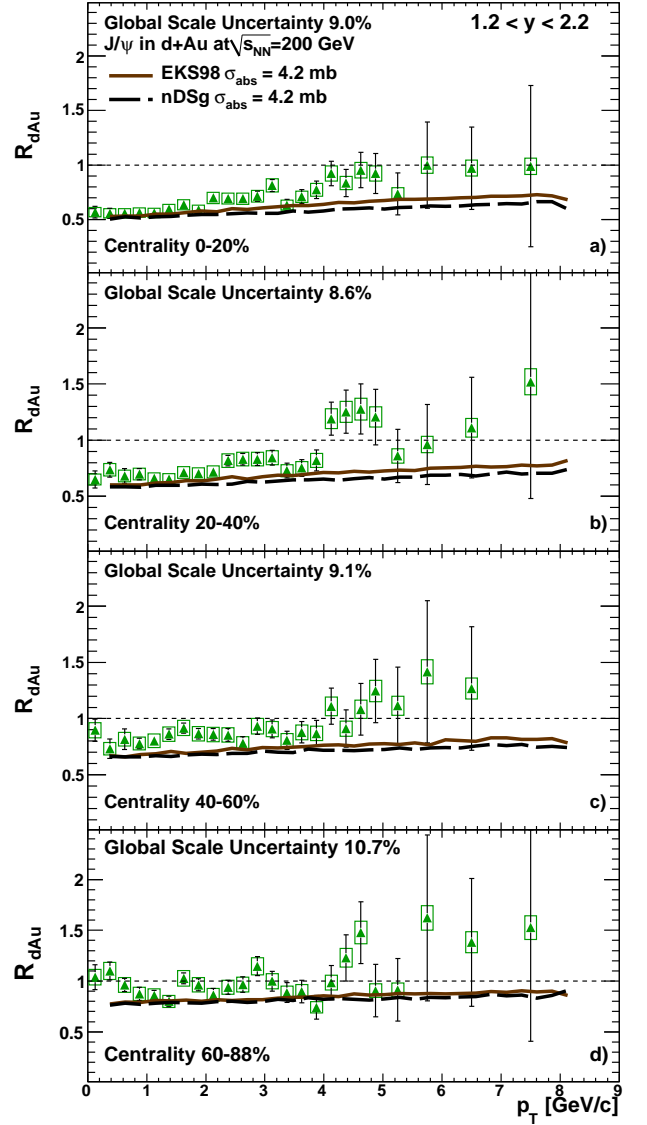}
\caption{(Color Online)
$J/\psi\rightarrow \mu^+\mu^-$ \rdau, as a 
function of \pt for a) central, b) midcentral, c) 
midperipheral, and d) peripheral events at \fory. Curves are calculations 
by Lansberg et al.~\cite{Ferreiro:2012zb} discussed in the text. }
\label{fig:rdauptcent_for}
\end{figure}

Figures \ref{fig:rdauptcent_bac}, \ref{fig:rdauptcent_mid}, and 
\ref{fig:rdauptcent_for} show \rdau vs \pt in four centrality bins for 
backward rapidity, midrapidity, and forward rapidity, respectively. 
Numerical values can be found in 
Tables~\ref{tab:rdaucent0},~\ref{tab:rdaucent1},~\ref{tab:rdaucent2} 
and~\ref{tab:rdaucent3} for 0--20\%, 20--40\%, 40--60\% and 60--88\% 
central collisions respectively. For peripheral collisions the \rdau 
remains consistent with 1.0 within statistical and systematic 
uncertainties across all \pt in all rapidity regions.

%%%%%%%%%%%%%%%%%%%%%%%%%%%%%%%%%%%%%%%%%%%%%%%%%%%%%%%% Fig_11

%%%%%%%%%%%%%%%%%%%%%%%%%%%%%%%%%%%%%%%%%%%%%%%%%%%%%%%% Fig_12

%%%%%%%%%%%%%%%%%%%%%%%%%%%%%%%%%%%%%%%%%%%%%%%%%%%%%%%% Fig_13

\subsection{Comparison with Model Predictions \label{sec:rdaumodel}}

As mentioned previously, various models have been suggested to describe 
the cold nuclear matter effects on \jpsi production. The models that will 
be discussed here include a combination of physical effects such as 
shadowing, nuclear breakup, and the Cronin effect.

Shadowing, the modification of the parton distributions within a nucleus, 
is calculated using parametrizations of deep inelastic scattering data in 
the form of nuclear modified parton distribution functions (nPDF's). 
There are a number of nPDF sets available, including 
nDSg~\cite{deFlorian:2003qf}, EKS98~\cite{Eskola:1998df} and 
EPS09~\cite{Eskola:2009uj}, which provide distributions of this 
modification based on different parametrizations of the available data. 
For \jpsi production in \dau collisions the relevant distributions are 
those providing the modification of the gluon distribution within a Au 
nucleus, as \jpsi's are produced primarily through gluon fusion at 
\sqsn=200 GeV. The nPDF's provide modifications as a function of parton 
momentum fraction ($x$) and energy transfer ($Q^2$). Knowledge of the 
\jpsi production kinematics is then needed to produce a modification to 
\jpsi production in \dau collisions. For \jpsi production at backward 
rapidity and $0<\pt<8$~\gevc, a range of roughly $0.051<x<0.39$ in the Au 
nucleus is probed, assuming simple 2$\rightarrow$1 kinematics. While 
2$\rightarrow$1 kinematics are inadequate to describe the production of a 
\jpsi with nonzero \pt, they are used here to provide a simple estimation 
of the $x$ and $Q^{2}$ ranges covered. Likewise, midrapidity covers 
roughly $0.0094<x<0.071$ and forward rapidity covers roughly 
$0.0017<x<0.013$. A range of roughly $10<Q^2[{\rm GeV}^2/c^2]<74$ is 
probed at each rapidity under the same assumptions. The data thus provide 
a strong constraint to shadowing models over a wide range of $x$ and 
$Q^2$.

Nuclear breakup, the dissociation of $c\bar{c}$ pairs that would have 
formed \jpsi's through collisions with nucleons, is often parametrized 
through a breakup cross section. Little theoretical or experimental 
guidance currently exists on the exact nature of this effect due to the 
many complications and competing effects involved in \jpsi production in 
$p(d)+A$ collisions. Often this effect is modeled by a simple 
``effective'' cross section, which remains constant with \pt, however 
there are a number of models, including a dynamic breakup cross section 
that changes based on the kinematics of the produced \jpsi.

The broadening of the \pt distribution, termed the Cronin 
effect~\cite{Cronin:1974zm}, is typically attributed to multiple elastic 
scattering of the incoming parton before the hard collision that produces 
the \jpsi. This modifies the \pt dependence of the \jpsi production by 
adding \pt vectorially to the incoming parton. This 
generally causes a decrease in \jpsi production at low \pt and a 
compensating increase at higher \pt ($\pt\approx 5-10$ \gevc), which 
eventually falls off at yet higher \pt ($\pt\approx 10$ \gevc).

The first set of model calculations that we discuss is by Kopeliovich et 
al.~\cite{Kopeliovich:2011zz, Kopeliovich:2010vk} calculates the effects 
on a $c\bar{c}$ dipole propagating through a nucleus. The \jpsi 
production is calculated based on 2$\rightarrow$1 kinematics,
\begin{equation}
x=\frac{\sqrt{\langle M_{c\bar{c}}^2\rangle+\langle\pt^2\rangle}}{\sqrt{s}}e^{-y},
\label{eq:kopx}
\end{equation}
where $\langle M_{c\bar{c}}^2\rangle=2M_{J/\psi}^2$ is fixed based on the 
$c\bar{c}$ invariant mass distribution predicted by the color singlet 
model. The calculation includes shadowing, taken from the nDSg nPDF set, 
as well as nuclear breakup and the Cronin effect. The nuclear breakup is 
calculated using a parametrization of the dipole cross section fitted to 
measurements of the proton structure function at 
HERA~\cite{GolecBiernat:1999qd}, yielding a breakup cross section that is 
dependent on kinematics of the \jpsi. The results from this calculation 
are shown for the 0--100\% \rdau at all rapidities in 
Fig.~\ref{fig:rdauptmb} as the dot-dashed curves. This is a parameter free 
calculation, with no overall normalization or fits to the data presented 
here. The \pt shape is in good agreement with the data at mid and forward 
rapidity, but the theory shows a greater overall level of suppression 
than is seen in the data. At backward rapidity there is a disagreement 
with the shape of the \pt distribution. While the theory predicts a 
similar \pt shape at all rapidities, the data show a much faster rise in 
\rdau with increasing \pt at backward rapidity. It is also worth noting, 
as shown in~\cite{Nagle:2010ix}, that this model cannot simultaneously 
describe the rapidity dependence of the PHENIX \rdau and \rcp, which is 
the ratio of \rdau in central collisions to the \rdau in peripheral 
collisions, for \jpsi production and therefore may not have an accurate 
description of the geometric dependence of the modification.

A second set of model calculations, performed by Lansberg et 
al.~\cite{Ferreiro:2008wc, Ferreiro:2012zb}, are shown in 
Figure~\ref{fig:rdauptmb}. This model uses a Monte-Carlo approach within 
a Glauber model of \dau collisions. The \jpsi production is calculated 
using the color singlet model that utilizes $2\rightarrow2$ kinematics, 
namely $g+g\rightarrow\jpsi+g$, where the majority of the \jpsi \pt is 
balanced by the emission of a hard gluon in the final state, rather than 
$2\rightarrow1$ processes, where the \jpsi \pt comes entirely from the 
transverse momentum carried by the colliding gluons. The \jpsi production 
is modified in \dau collisions by shadowing effects parametrized using 
various nPDF sets. The calculations shown in Fig.~\ref{fig:rdauptmb} 
utilize the nDSg nPDF set. Similar calculations using the EKS98 and 
EPS08~\cite{Eskola:2008ca} nPDF sets can be found 
in~\cite{Ferreiro:2012zb}. Nuclear breakup of the \jpsi is taken into 
account through the use of an effective, \pt-independent, absorption 
cross section of 4.2 mb. Results using $\sigma_{abs}=$0, 2.6, and 6 mb 
can also be found in~\cite{Ferreiro:2012zb}. We have chosen to highlight 
only $\sigma_{abs}=4.2$ mb here as it reproduces the rapidity dependence 
of the 60--88\% \rdau reasonably well~\cite{Ferreiro:2012zb} where 
shadowing corrections are expected to be small. The results of this 
calculation, shown in Fig.~\ref{fig:rdauptmb} for 0--100\% \rdau at all 
rapidities, shows reasonable agreement with the overall level of 
modification seen at low \pt in the data at mid and forward rapidities 
while the calculation predicts a flatter distribution with increasing \pt 
than is seen in the data. The shape of the distribution at backward 
rapidity is markedly different than the data. While the data rapidly 
increase to $\rdau\approx 1$ at low \pt, the calculation shows a \rdau 
that is essentially constant with increasing \pt.

When comparing the two sets of model calculations in Fig.~\ref{fig:rdauptmb} 
the calculations from 
Kopeliovich et al.~\cite{Kopeliovich:2011zz, Kopeliovich:2010vk} 
have a different and more pronounced shape when compared to the calculations 
from Lansberg, et al~\cite{Ferreiro:2008wc, Ferreiro:2012zb}.  Both sets 
of calculations utilize the nDSg 
nPDF set, suggesting a common contribution from shadowing. However, the 
\jpsi production kinematics are calculated differently, which will lead to 
some difference in the shadowing contribution. The calculations from 
Kopeliovich et al. include the Cronin effect, which provides a decrease 
in \jpsi production at low \pt and an increase at higher \pt, creating an 
\rdau that exhibits less suppression at high \pt than at low \pt. The 
calculations from Lansberg et al. do not include the Cronin effect, and 
therefore the \pt shape of \rdau should be dominated by the effect of 
shadowing, and therefore the choice of nPDF set.

The spatial dependence of the shadowing has been taken into account 
in~\cite{Ferreiro:2012zb}, where it is assumed that the shadowing is 
proportional to the local density. This assumption allows for calculation 
of the \rdau vs \pt in different centrality bins. The results of the 
calculation in the four PHENIX centrality bins are shown in 
Figures~\ref{fig:rdauptcent_bac},~\ref{fig:rdauptcent_mid} 
and~\ref{fig:rdauptcent_for} for backward, mid, and forward rapidity, 
respectively. Here we have chosen to include calculations using the EKS98 
nPDF set along with those using the nDSg nPDF set as this will provide a 
direct comparison between the effects due to different nPDF sets, since 
the \jpsi production kinematics and $\sigma_{abs}$ values are identical 
between the two calculations. At mid and forward rapidity the 
calculations are similar to each other and show reasonable agreement with 
the \rdau distributions within the current statistical and systematic 
uncertainties, although the calculation appears to predict a slightly 
larger average suppression for peripheral collisions at forward rapidity 
than is seen in the data. This could be due to the value of 
$\sigma_{abs}$ used at forward rapidity, as the value of 4.2 mb was 
chosen by eye rather than fitted to the data, and it may not be 
independent of $y$.

At backward rapidity the calculations are in disagreement with the data 
for all but the most peripheral collisions, where both the calculations 
and the data show an \rdau consistent with 1.0 at all \pt. While the 
calculations at backward rapidity using the nDSg nPDF set are roughly 
constant with \pt for each centrality, the calculations using the EKS98 
nPDF set show an enhancement in the suppression of \rdau with increasing 
\pt for central and midcentral collisions, whereas the data shows the 
opposite trend. At backward rapidity and low \pt (Bjorken $x\approx 0.1$ 
for the parton in the Au nucleus) production occurs in the anti-shadowing 
region, while at high \pt ($x\approx0.3$) production begins to move 
towards what is termed the EMC~\cite{Aubert:1983xm} region. In 1986, a 
suppression of the quark distributions within nuclei was discovered in 
the range $0.35<x<0.7$ by the European Muon Collaboration 
(EMC)~\cite{Aubert:1983xm} in deep inelastic scattering. While there is 
still debate about the source of this suppression in the quark 
distributions, no direct evidence of an EMC effect has yet been reported 
in the gluon distributions. Few constraints exist in this region, and 
there is large disagreement in the modification of the gluon density 
between nPDF's. The nDSg nPDF set includes no suppression in the EMC 
region, and only a small anti-shadowing effect, while the EKS98 nPDF 
exhibits a suppression in the EMC region similar to that observed in the 
quark distributions, and a larger anti-shadowing effect 
(see~\cite{Eskola:2009uj} for a comparison of nPDF sets). The larger 
anti-shadowing combined with the inclusion of an EMC effect in the EKS98 
nPDF set cause a decrease in the calculated \rdau as \pt (and 
correspondingly, $x$) increases. The lack of a strong anti-shadowing 
effect combined with the absence of an EMC effect in the nDSg nPDF causes 
the calculation of \rdau to remain roughly constant with increasing \pt.

In~\cite{Ferreiro:2011xy} the authors infer from measurements of 
$\Upsilon$ production at RHIC that a strong EMC effect must be present to 
explain the observed modification. Depending on the mapping of the \jpsi 
$y$ and \pt to $x$, which is model dependent, the high \pt data at 
backward rapidity may allow us to probe this region. The large 
uncertainties present in the high \pt \rdau, along with complications 
from competing physics effects in this region, however, prevent any 
strong conclusions from being drawn at this time.

%%%%%%%%%%%%%%%%%%%%%%%%%%%%%%%%%%%%%%%%%%%%%%%%%%%%%%%% Fig_14
\begin{figure}[!th]
\includegraphics[width=1.0\linewidth]{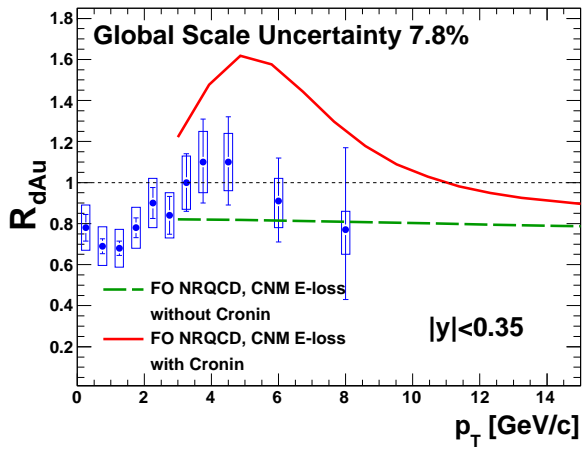}
\caption{(Color Online)
\jpsi \rdau, as a function of \pt momentum for midrapidity 0--100\% 
centrality integrated \dau collisions. The curves are theoretical 
calculations from~\cite{Sharma:2012dy} described in the text.}
\label{fig:rdauptmid_vitev}
\end{figure}

A third set of model calculations by Sharma and 
Vitev~\cite{Sharma:2012dy} is compared with the midrapidity 0--100\% 
centrality integrated \rdau in Fig.~\ref{fig:rdauptmid_vitev}. This 
model describes \jpsi production using nonrelativistic quantum 
chromodynamics (NRQCD). The effect of nuclear shadowing is calculated 
using EKS98 in the EMC region ($x>0.25$), while for lower values, power 
suppressed coherent final-state scattering leads to a modification of 
parton $x$~\cite{Vitev:2006bi}. Initial state energy loss, which accounts 
for the radiative energy loss of the incoming particles through multiple 
interactions with the target nucleus is included. This effect reduces the 
energy of the incoming parton, so, to achieve the same final-state 
kinematics the parton must have a greater momentum, and therefore a 
larger value of $x$. This effectively shifts the portion of the gluon 
distribution sampled to higher $x$. Also included is a calculation of the 
Cronin effect. The solid curve in Fig.~\ref{fig:rdauptmid_vitev} shows 
the full calculation including the Cronin effect. The dashed curve in 
Figure~\ref{fig:rdauptmid_vitev} is the same calculation without the 
Cronin effect. This comparison gives a direct indication of the 
contribution from the Cronin effect, which is evidently over predicted 
when compared to the data. The results presented here will hopefully 
provide a much needed constraint on the Cronin effect at RHIC energies. 
The calculation including the Cronin effect indicate an \rdau that 
decreases at higher \pt. This is consistent with data, however the 
current statistical and systematic uncertainties make determining the 
precise trend of \rdau difficult at high \pt. Better data with a larger 
\pt coverage is needed to determine the \jpsi modification at higher \pt.

%%========================================================================%%
%%========================================================================%%
%%========================================================================%%
\section{Summary \& Conclusions \label{sec:sum}}

We have measured the \jpsi invariant yield and $R_{d{\rm Au}}$, 
as a function of \pt over three rapidity ranges in \dau collisions 
at $\sqsn=200$ GeV using the PHENIX detector. These measurements provide 
a large improvement in statistical precision and \pt reach over the 
previously published PHENIX \dau 
results~\cite{Adare:2007gn,Adler:2005ph}, and are the first measurements 
of the centrality dependence of the \jpsi \pt distribution in \dau 
collisions by PHENIX. The $\Delta\mptsq$ values determined from the data 
show a marked increase with \Ncoll that is similar at all rapidities.

The \rdau vs \pt displays similar behavior at mid and forward rapidity, 
showing suppression at low \pt with a gradual increase to a value 
consistent with 1.0. The \rdau at backward rapidity has a different 
distribution with \pt, showing a more rapid increase from suppression to 
a value of 1.0, and transitioning to $\rdau>1.0$ above 2 \gevc. These 
trends are greater for central collisions, while the peripheral 
collisions show \rdau consistent with 1.0 across all rapidities.

We find an average \rdau for $\pt>4$ \gevc of $1.27\pm0.06\pm0.11$ at 
backward rapidity, and an \rdau consistent with 1.0 at mid and forward 
rapidity. This implies a CNM contribution in $A+A$ collisions that is 
likely consistent with 1.0 at high \pt across all rapidity. This could 
potentially explain the reported increase in $R_{\rm AA}$ with increasing 
\pt~\cite{Abelev:2009qaa}. However more data and further work to 
understand the propagation of \rdau to $R_{\rm AA}$ is needed to confirm 
this.

A comparison of the measured \rdau with three types of theoretical 
calculations was shown. The parameter independent dipole model of \jpsi 
production in $p+A$ collisions agrees well with the shape of the data at 
mid and forward rapidities, while the shape of the predicted \pt 
dependence is different from the data at backward rapidity. However the 
suppression is over-predicted at all rapidities. The second model uses 
$2\rightarrow2$ \jpsi production kinematics coupled with shadowing taken 
from both EKS98 and nDSg nPDF sets as well as an effective absorption 
cross section of 4.2 mb. The calculations with both EKS98 and nDSg show 
good agreement with the data at midrapidity in each centrality bin, as 
well as the centrality integrated case. At forward rapidity the shape of 
the distribution is in reasonable agreement with the data, while the 
overall level of suppression seems to be greater in the model 
calculations than the data. At backward rapidity, the model calculations 
using both EKS98 and nDSG nPDF sets are in strong disagreement with the 
data. At backward rapidity calculations using the nDSg nPDF set show a 
suppression that is constant with \pt, while those using the EKS98 nPDF 
set predict an increase of suppression with increasing \pt. The data show 
the opposite trend. The third model, an NRQCD calculation of high \pt 
\jpsi production show a Cronin effect, which although generally 
consistent with the data, is significantly larger than observed in the 
data, and a suppression at high \pt that cannot be confirmed due to the 
large uncertainties at high \pt and the limited \pt reach of the current 
data.

In summary, the data presented here cover a large range in $x$ and $Q^2$, 
providing a further constraint on the modification of the gluon 
distribution in nuclei, as well as providing constraints on the size of 
the Cronin effect on \jpsi production at RHIC.

%%========================================================================%%
%%========================================================================%%
%%========================================================================%%

%%%%%%%%%%%%%%%%%%%%%%%%%  Acknowledgements 

\section*{ACKNOWLEDGMENTS}   % Run-8 long from for PRC, PLB, etc.

We thank the staff of the Collider-Accelerator and Physics
Departments at Brookhaven National Laboratory and the staff of
the other PHENIX participating institutions for their vital
contributions.  
We thank Jean-Philippe Lansberg, Nicolas Matagne, Boris 
Kopeliovich, Ivan Vitev, and Rishi Sharma for useful discussions and 
theoretical calculations.
We acknowledge support from the 
Office of Nuclear Physics in the
Office of Science of the Department of Energy, the
National Science Foundation, Abilene Christian University
Research Council, Research Foundation of SUNY, and Dean of the
College of Arts and Sciences, Vanderbilt University (U.S.A),
Ministry of Education, Culture, Sports, Science, and Technology
and the Japan Society for the Promotion of Science (Japan),
Conselho Nacional de Desenvolvimento Cient\'{\i}fico e
Tecnol{\'o}gico and Funda\c c{\~a}o de Amparo {\`a} Pesquisa do
Estado de S{\~a}o Paulo (Brazil),
Natural Science Foundation of China (P.~R.~China),
Ministry of Education, Youth and Sports (Czech Republic),
Centre National de la Recherche Scientifique, Commissariat
{\`a} l'{\'E}nergie Atomique, and Institut National de Physique
Nucl{\'e}aire et de Physique des Particules (France),
Ministry of Industry, Science and Tekhnologies,
Bundesministerium f\"ur Bildung und Forschung, Deutscher
Akademischer Austausch Dienst, and Alexander von Humboldt Stiftung (Germany),
Hungarian National Science Fund, OTKA (Hungary), 
Department of Atomic Energy and Department of Science and Technology (India), 
Israel Science Foundation (Israel), 
National Research Foundation and WCU program of the 
Ministry Education Science and Technology (Korea),
Ministry of Education and Science, Russian Academy of Sciences,
Federal Agency of Atomic Energy (Russia),
VR and the Wallenberg Foundation (Sweden), 
the U.S. Civilian Research and Development Foundation for the
Independent States of the Former Soviet Union, 
the US-Hungarian Fulbright Foundation for Educational Exchange,
and the US-Israel Binational Science Foundation.

\appendix

\section{Details on the calculation of \mptsq \label{sec:detailmptsq}}

\subsection{Fitting The \jpsi Invariant Yields \label{sec:ptfits}}

\begin{figure*}[thb]
\includegraphics[width=0.3\linewidth]{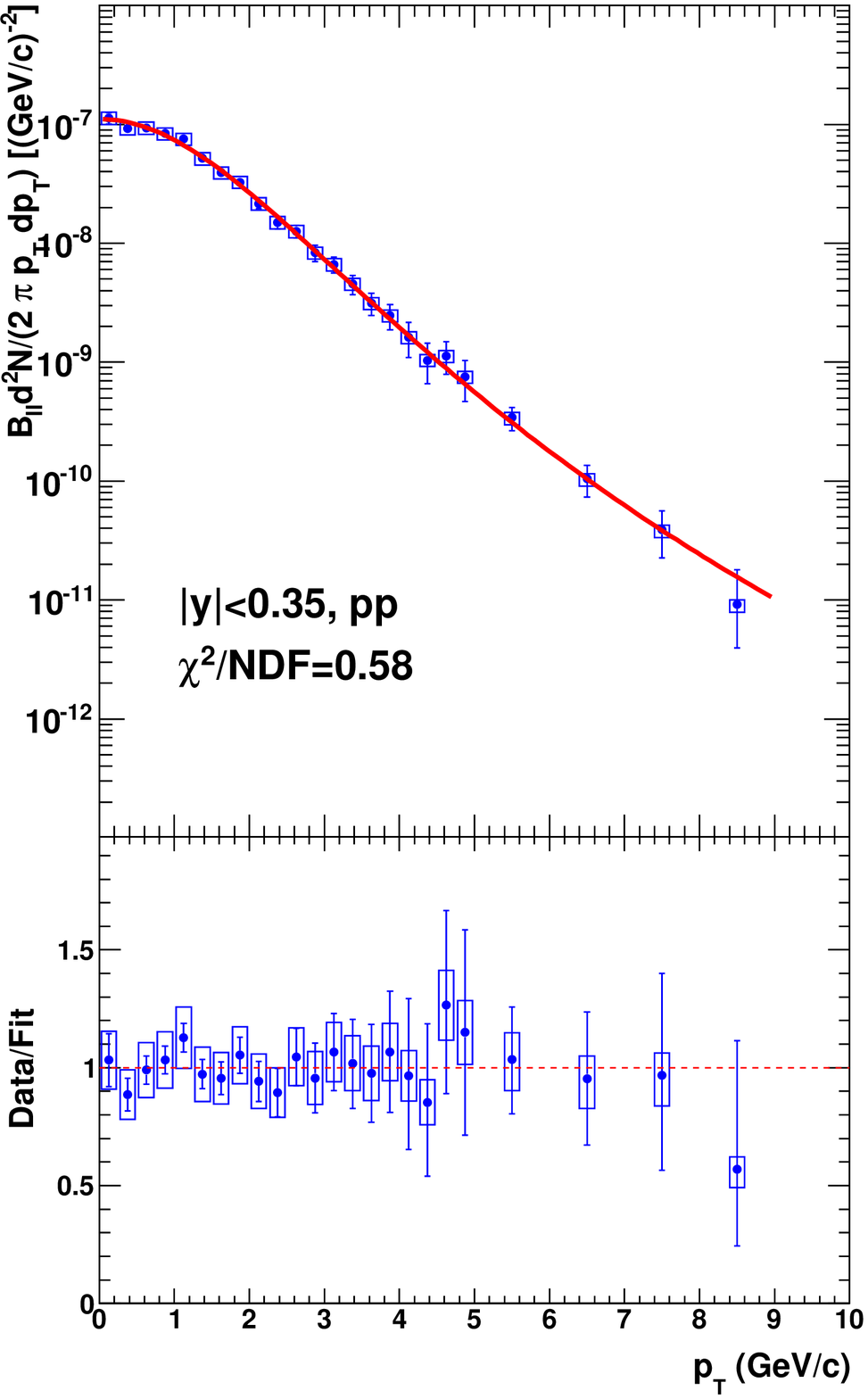}  
\includegraphics[width=0.3\linewidth]{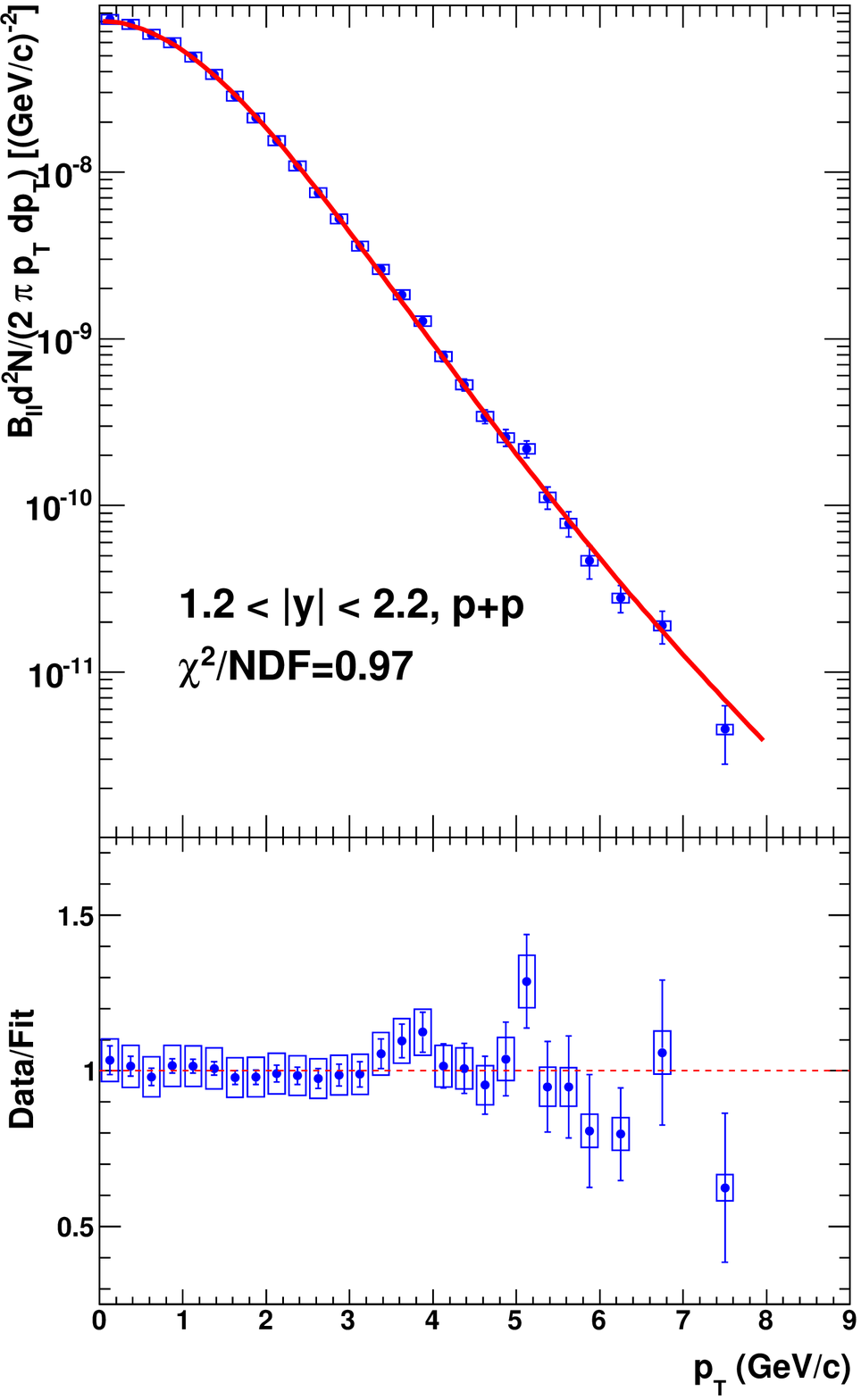}  
\caption{(Color Online)
Results of modified Kaplan fits to midrapidity \pp (Left) and forward 
rapidity \pp (Right).
}
\label{fig:ptfitpp}
%\end{figure*}

%%%%%%%%%%%%%%%%%%%%%%%%%%%%%%%%%%%%%%%%%%%%%%%%%%%%%%  Fig_16
%\begin{figure*}[thb]
\includegraphics[width=0.3\linewidth]{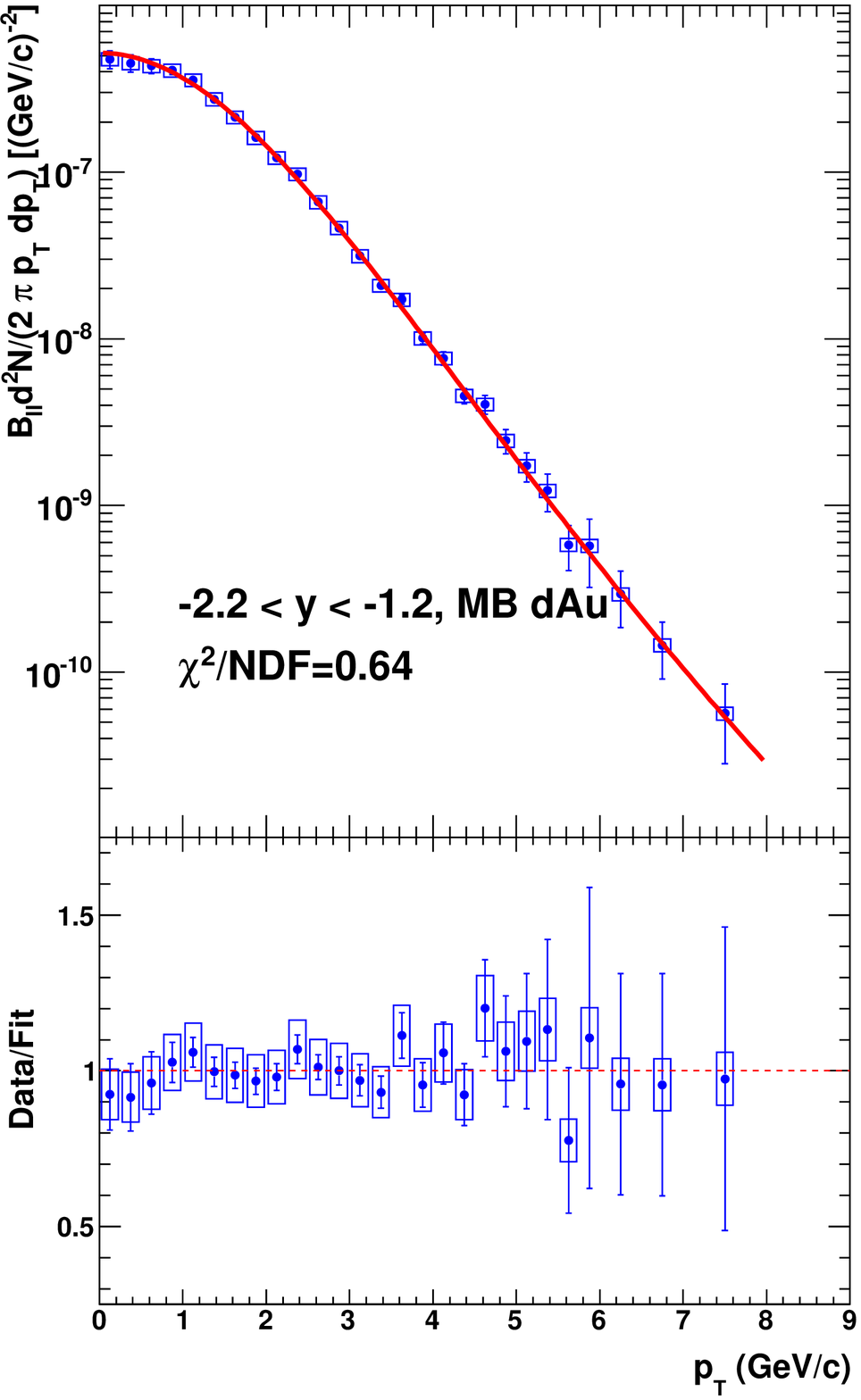}  
\includegraphics[width=0.3\linewidth]{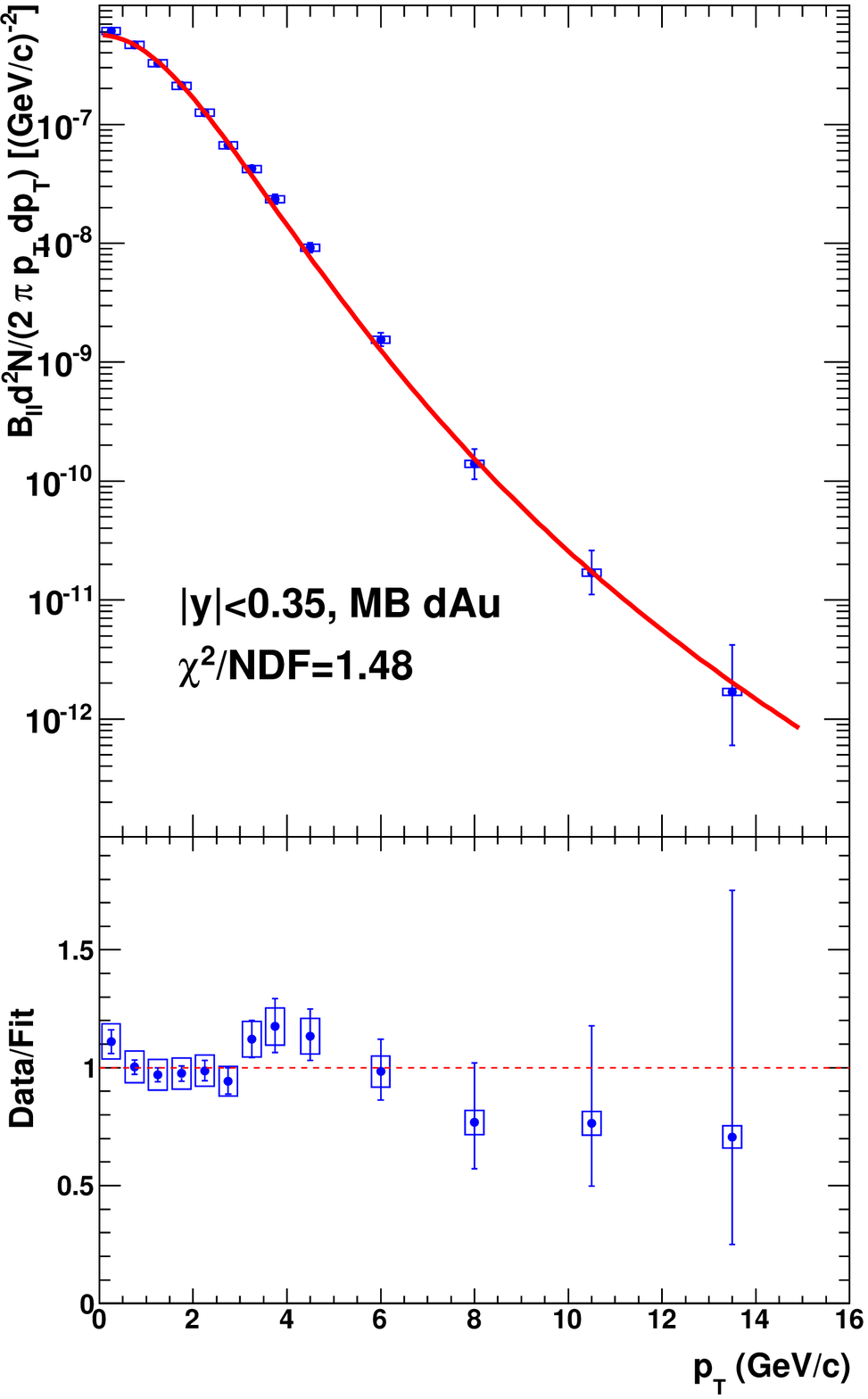}  
\includegraphics[width=0.3\linewidth]{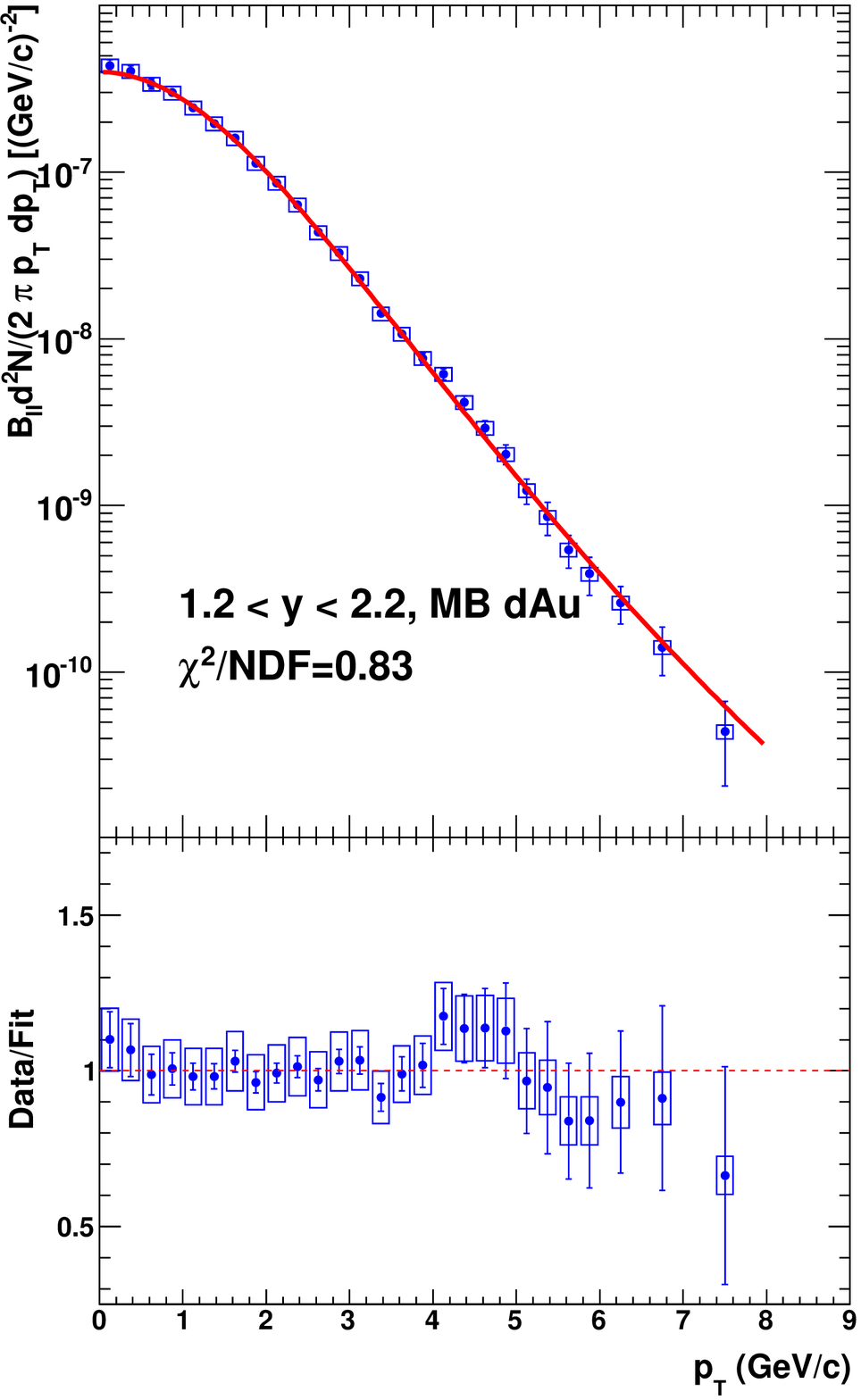}  
\caption{(Color Online)
Results of modified Kaplan fits to backward rapidity 0--100\% \dau 
(Left), midrapidity 0-100\% \dau (Center), and forward rapidity 0-100\% 
\dau (Right).
}
\label{fig:ptfitMBdau}
\end{figure*}

\begin{figure*}[thb]
\includegraphics[width=0.8\linewidth]{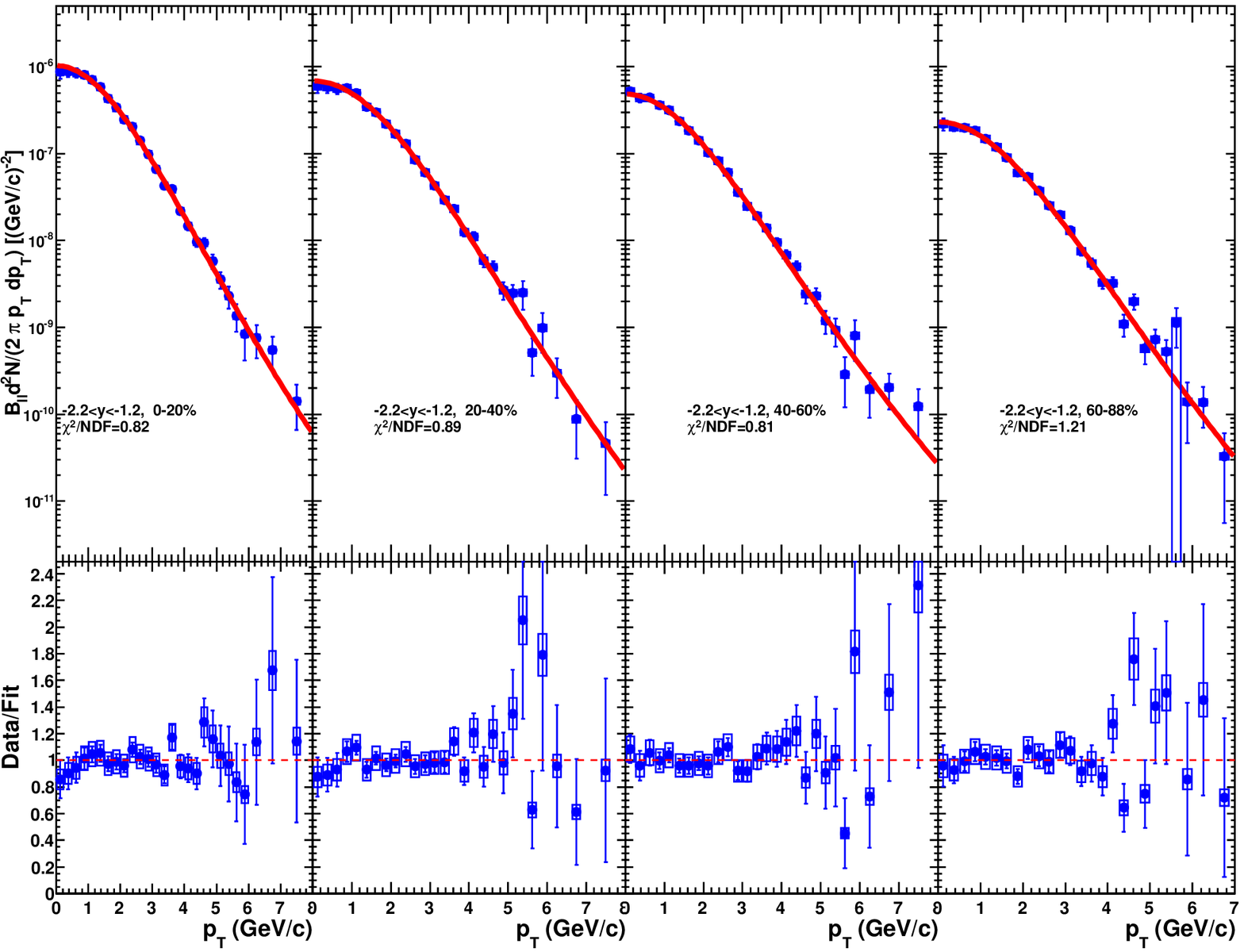}  
\caption{(Color Online)
Results of modified Kaplan fits to backward rapidity \dau collisions for 
each centrality.
}
\label{fig:ptfitcentbac}
%\end{figure*}

%%%%%%%%%%%%%%%%%%%%%%%%%%%%%%%%%%%%%%%%%%%%%%%%%%%%%%  Fig_18
%\begin{figure*}[thb]
\includegraphics[width=0.8\linewidth]{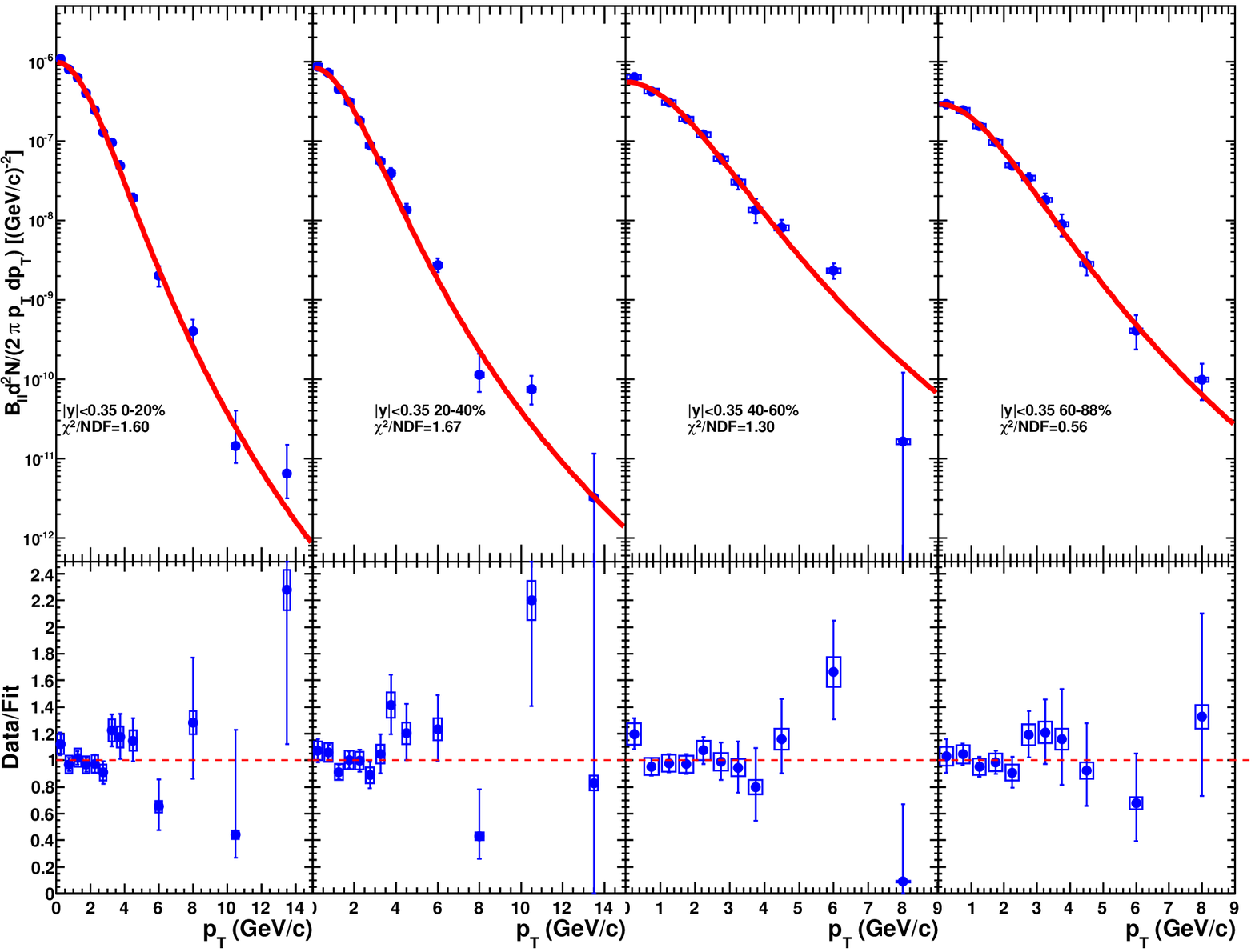}  
\caption{(Color Online)
Results of modified Kaplan fits to midrapidity \dau collisions for each 
centrality.
}
\label{fig:ptfitcentmid}
\end{figure*}

\begin{figure*}[thb]
\includegraphics[width=0.8\linewidth]{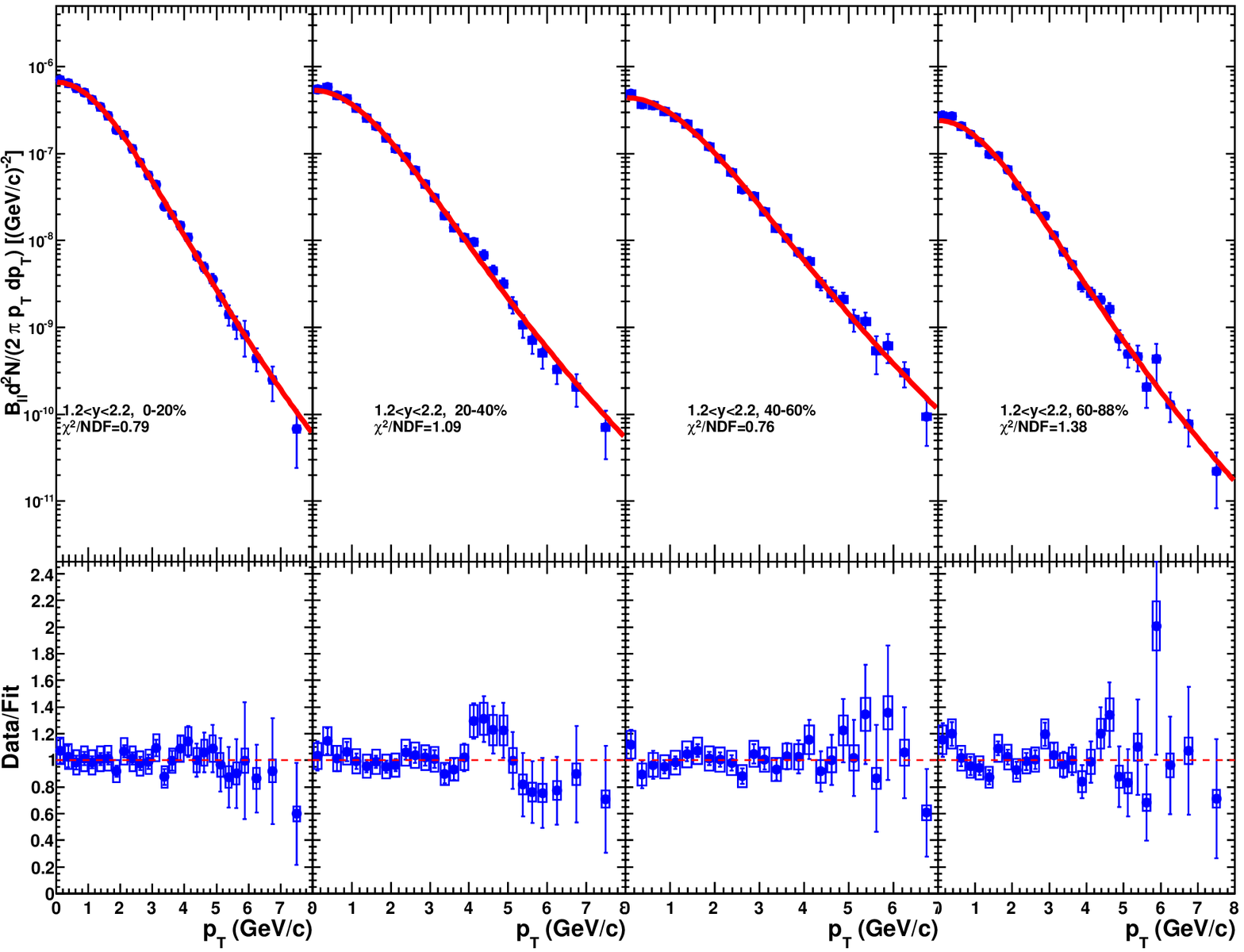}  
\caption{(Color Online)
Results of modified Kaplan fits to forward rapidity \dau collisions for 
each centrality.
}
\label{fig:ptfitcentfor}
\end{figure*}

The \jpsi invariant yields as a function of \pt were fitted with a 
modified Kaplan function of the form
\begin{equation}
f(\pt)=p_0\left(1-\left(\frac{\pt}{p_1}\right)^2\right)^{p_2}.
\label{eq:modkaplan2}
\end{equation}
The data points were compared to the integral of the function over the 
\pt bin when calculating the $\chi^2$. The fit results, along with the 
ratio of the data to the fit are shown in Fig.~\ref{fig:ptfitpp} for 
\pp collisions and Figure~\ref{fig:ptfitMBdau} for 0--100\% centrality 
integrated \dau collisions. The fit results for each centrality bin are 
shown in Figs.~\ref{fig:ptfitcentbac},~\ref{fig:ptfitcentmid}, 
and~\ref{fig:ptfitcentfor} for backward, mid, and forward rapidities, 
respectively.

\subsection{Calculating the Correction Factor $k$}

To account for the fact that the experimental upper \pt limits on the 
\jpsi invariant yield distributions vary with rapidity and centrality, a 
correction factor was calculated using the fits described in 
Sec.~\ref{sec:ptfits}. The ratio
\begin{equation} 
k=\frac{\mptsq[0,\infty]}{\mptsq[0,\pt^{\rm max}]}
\label{eq:mptsq_corr2}
\end{equation}
was calculated from the fit and applied to the numerically calculated 
\mptsqmax. The correction factors are shown in Table~\ref{tab:kfactor}, 
and are in all cases small ($k<1.03$). The uncertainty on $k$ is derived 
from the fit uncertainty by varying the data points within their 
statistical uncertainties, refitting, and thereby finding the variation 
in $k$.

%%%%%%%%%%%%%%%%%%%%%%%%%%%%%%%%%%%%%%%%%%%%%%%%%%%%%%  Fig_15

%%%%%%%%%%%%%%%%%%%%%%%%%%%%%%%%%%%%%%%%%%%%%%%%%%%%%  Fig_17

%%%%%%%%%%%%%%%%%%%%%%%%%%%%%%%%%%%%%%%%%%%%%%%%%%%%%%  Fig_19

%====================================================== Table_VI
\begin{table*}[tbh]
\caption{\label{tab:kfactor} \mptsq correction factors, $k$, for \pp and 
\dau collisions.}
\begin{ruledtabular}\begin{tabular}{cccc}
System & y range & Centrality & $k$ \\
\hline
\pp  & \muony &         & 1.006$\pm$0.001 \\
\pp  & \midy  &         & 1.018$\pm$0.007 \\
\\
\dau & \bacy  & 0--100\% & 1.005$\pm$0.001 \\
\dau & \midy  & 0--100\% & 1.002$\pm$0.001 \\
\dau & \fory  & 0--100\% & 1.010$\pm$0.002 \\
\\
\dau & \bacy &  0--20\% & 1.005$\pm$0.002 \\
\dau & \bacy & 20--40\% & 1.002$\pm$0.001 \\
\dau & \bacy & 40--60\% & 1.006$\pm$0.002 \\
\dau & \bacy & 60--88\% & 1.008$\pm$0.003 \\
\\
\dau & \midy &  0--20\% & 1.001$\pm$0.001 \\
\dau & \midy & 20--40\% & 1.002$\pm$0.002 \\
\dau & \midy & 40--60\% & 1.024$\pm$0.017 \\
\dau & \midy & 60--88\% & 1.020$\pm$0.024 \\
\\
\dau & \fory &  0--20\% & 1.009$\pm$0.003 \\
\dau & \fory & 20--40\% & 1.012$\pm$0.003 \\
\dau & \fory & 40--60\% & 1.023$\pm$0.006 \\
\dau & \fory & 60--88\% & 1.010$\pm$0.004 \\
\end{tabular}\end{ruledtabular}
\end{table*}

\subsection{Propagating the Type B uncertainties to \mptsq}
\label{sec:typeBmptsq}

When propagating the Type B systematic uncertainties on the \jpsi 
invariant yields to the calculated \mptsq values, the type B 
uncertainties are assumed to be normally distributed. With this 
assumption we independently sample the uncertainty distribution of the 
first and the last data point of the \pt distribution. We then assume the 
Type B uncertainties are linearly correlated between these two values. 
The resulting distribution of the \mptsq values that arises from this 
procedure gives an estimate of the effect of the Type B uncertainties on 
the value of \mptsq.

The Type C systematic uncertainties on the \jpsi invariant yields do not 
affect the calculation of \mptsq. The Type C uncertainties are a global 
uncertainty, which cancels in the calculation.

\clearpage

\section{Data Tables \label{sec:datatables}}

%====================================================== Table_VII
\begin{table}[!ht]
\caption{\label{tab:mbrdaubac}
Data tables for 0--100\% centrality-integrated \rdau at \bacy.}
\begin{ruledtabular}\begin{tabular}{ccccc}
\pt [\gevc] & \rdau & Type A & Type B & Type C \\
0.00 - 0.25 & 0.759 & $\pm$0.1 & $\pm$0.053 & $\pm$0.063 \\ 
0.25 - 0.50 & 0.772 & $\pm$0.094 & $\pm$0.054 & $\pm$0.064 \\ 
0.50 - 0.75 & 0.853 & $\pm$0.092 & $\pm$0.059 & $\pm$0.071 \\ 
0.75 - 1.00 & 0.899 & $\pm$0.06 & $\pm$0.062 & $\pm$0.074 \\ 
1.00 - 1.25 & 0.955 & $\pm$0.048 & $\pm$0.065 & $\pm$0.079 \\ 
1.25 - 1.50 & 0.934 & $\pm$0.048 & $\pm$0.064 & $\pm$0.077 \\ 
1.50 - 1.75 & 0.988 & $\pm$0.048 & $\pm$0.068 & $\pm$0.082 \\ 
1.75 - 2.00 & 1.000 & $\pm$0.049 & $\pm$0.069 & $\pm$0.083 \\ 
2.00 - 2.25 & 1.043 & $\pm$0.054 & $\pm$0.072 & $\pm$0.086 \\ 
2.25 - 2.50 & 1.182 & $\pm$0.061 & $\pm$0.081 & $\pm$0.098 \\ 
2.50 - 2.75 & 1.159 & $\pm$0.059 & $\pm$0.08 & $\pm$0.096 \\ 
2.75 - 3.00 & 1.161 & $\pm$0.067 & $\pm$0.08 & $\pm$0.096 \\ 
3.00 - 3.25 & 1.150 & $\pm$0.075 & $\pm$0.079 & $\pm$0.095 \\ 
3.25 - 3.50 & 1.059 & $\pm$0.076 & $\pm$0.073 & $\pm$0.088 \\ 
3.50 - 3.75 & 1.234 & $\pm$0.101 & $\pm$0.085 & $\pm$0.102 \\ 
3.75 - 4.00 & 1.043 & $\pm$0.098 & $\pm$0.072 & $\pm$0.086 \\ 
4.00 - 4.25 & 1.285 & $\pm$0.15 & $\pm$0.089 & $\pm$0.106 \\ 
4.25 - 4.50 & 1.133 & $\pm$0.152 & $\pm$0.078 & $\pm$0.094 \\ 
4.50 - 4.75 & 1.556 & $\pm$0.252 & $\pm$0.108 & $\pm$0.129 \\ 
4.75 - 5.00 & 1.265 & $\pm$0.256 & $\pm$0.089 & $\pm$0.105 \\ 
5.00 - 5.50 & 1.186 & $\pm$0.303 & $\pm$0.083 & $\pm$0.098 \\ 
5.50 - 6.00 & 1.227 & $\pm$0.518 & $\pm$0.085 & $\pm$0.101 \\ 
6.00 - 7.00 & 1.228 & $\pm$0.537 & $\pm$0.085 & $\pm$0.102 \\ 
7.00 - 8.00 & 1.643 & $\pm$1.036 & $\pm$0.116 & $\pm$0.136 \\ 
\end{tabular}\end{ruledtabular}
\end{table}

%====================================================== Table_VIII
\begin{table}[!ht]
\caption{\label{tab:mbrdaumid}
 Data tables for 0--100\% centrality-integrated \rdau at \midy.}
\begin{ruledtabular}\begin{tabular}{ccccc}
\pt [\gevc] & \rdau & Type A & Type B & Type C \\
0.0 - 0.5 & 0.78 & $\pm$0.065 & $\pm$0.11 & $\pm$0.061 \\ 
0.5 - 1.0 & 0.69 & $\pm$0.036 & $\pm$0.094 & $\pm$0.055 \\ 
1.0 - 1.5 & 0.68 & $\pm$0.035 & $\pm$0.092 & $\pm$0.053 \\ 
1.5 - 2.0 & 0.78 & $\pm$0.048 & $\pm$0.1 & $\pm$0.061 \\ 
2.0 - 2.5 & 0.90 & $\pm$0.076 & $\pm$0.12 & $\pm$0.071 \\ 
2.5 - 3.0 & 0.84 & $\pm$0.092 & $\pm$0.11 & $\pm$0.066 \\ 
3.0 - 3.5 & 1.00 & $\pm$0.14 & $\pm$0.13 & $\pm$0.078 \\ 
3.5 - 4.0 & 1.10 & $\pm$0.21 & $\pm$0.15 & $\pm$0.087 \\ 
4.0 - 5.0 & 1.10 & $^{+0.22}_{-0.21}$ & $\pm$0.14 & $\pm$0.084 \\ 
5.0 - 7.0 & 0.91 & $\pm$0.21 & $^{+0.11}_{-0.13}$ & $\pm$0.072 \\ 
7.0 - 9.0 & 0.77 & $^{+0.4}_{-0.34}$ & $^{+0.09}_{-0.12}$ & $\pm$0.06 \\ 
\end{tabular}\end{ruledtabular}
\end{table}

%====================================================== Table_IX
\begin{table}[!ht]
\caption{\label{tab:mbrdaufor}
 Data tables for 0--100\% centrality-integrated \rdau at \fory.}
\begin{ruledtabular}\begin{tabular}{ccccc}
\pt [\gevc] & \rdau & Type A & Type B & Type C \\
0.00 - 0.25 & 0.693 & $\pm$0.064 & $\pm$0.052 & $\pm$0.057 \\ 
0.25 - 0.50 & 0.690 & $\pm$0.059 & $\pm$0.052 & $\pm$0.057 \\ 
0.50 - 0.75 & 0.664 & $\pm$0.048 & $\pm$0.05 & $\pm$0.055 \\ 
0.75 - 1.00 & 0.659 & $\pm$0.037 & $\pm$0.049 & $\pm$0.054 \\ 
1.00 - 1.25 & 0.652 & $\pm$0.032 & $\pm$0.048 & $\pm$0.054 \\ 
1.25 - 1.50 & 0.671 & $\pm$0.032 & $\pm$0.05 & $\pm$0.055 \\ 
1.50 - 1.75 & 0.739 & $\pm$0.029 & $\pm$0.055 & $\pm$0.061 \\ 
1.75 - 2.00 & 0.703 & $\pm$0.029 & $\pm$0.052 & $\pm$0.058 \\ 
2.00 - 2.25 & 0.732 & $\pm$0.031 & $\pm$0.054 & $\pm$0.061 \\ 
2.25 - 2.50 & 0.772 & $\pm$0.035 & $\pm$0.057 & $\pm$0.064 \\ 
2.50 - 2.75 & 0.764 & $\pm$0.037 & $\pm$0.057 & $\pm$0.063 \\ 
2.75 - 3.00 & 0.821 & $\pm$0.043 & $\pm$0.061 & $\pm$0.068 \\ 
3.00 - 3.25 & 0.844 & $\pm$0.049 & $\pm$0.063 & $\pm$0.07 \\ 
3.25 - 3.50 & 0.716 & $\pm$0.048 & $\pm$0.053 & $\pm$0.059 \\ 
3.50 - 3.75 & 0.765 & $\pm$0.056 & $\pm$0.057 & $\pm$0.063 \\ 
3.75 - 4.00 & 0.786 & $\pm$0.07 & $\pm$0.059 & $\pm$0.065 \\ 
4.00 - 4.25 & 1.032 & $\pm$0.107 & $\pm$0.077 & $\pm$0.085 \\ 
4.25 - 4.50 & 1.030 & $\pm$0.129 & $\pm$0.077 & $\pm$0.085 \\ 
4.50 - 4.75 & 1.118 & $\pm$0.166 & $\pm$0.084 & $\pm$0.092 \\ 
4.75 - 5.00 & 1.047 & $\pm$0.186 & $\pm$0.079 & $\pm$0.087 \\ 
5.00 - 5.50 & 0.836 & $\pm$0.195 & $\pm$0.063 & $\pm$0.069 \\ 
5.50 - 6.00 & 0.984 & $\pm$0.298 & $\pm$0.074 & $\pm$0.081 \\ 
6.00 - 7.00 & 1.122 & $\pm$0.387 & $\pm$0.084 & $\pm$0.093 \\ 
7.00 - 8.00 & 1.275 & $\pm$0.83 & $\pm$0.097 & $\pm$0.105 \\ 
\end{tabular}\end{ruledtabular}
\end{table}

\begingroup \squeezetable

%====================================================== Table_X
\begin{table}[!ht]
\caption{\label{tab:rdaucent0}
 Data tables for \rdau as a function of \pt for 0--20\% centrality.}
%\tiny
\begin{ruledtabular}\begin{tabular}{cccccc}
y & \pt [\gevc] & \rdau & Type A & Type B & Type C \\
\hline
\bacy & 0.00 - 0.25 & 0.702 & $\pm$0.121 & $\pm$0.049 & $\pm$0.063 \\ 
\bacy & 0.25 - 0.50 & 0.760 & $\pm$0.104 & $\pm$0.053 & $\pm$0.068 \\ 
\bacy & 0.50 - 0.75 & 0.840 & $\pm$0.105 & $\pm$0.058 & $\pm$0.075 \\ 
\bacy & 0.75 - 1.00 & 0.894 & $\pm$0.076 & $\pm$0.062 & $\pm$0.08 \\ 
\bacy & 1.00 - 1.25 & 0.954 & $\pm$0.059 & $\pm$0.065 & $\pm$0.085 \\ 
\bacy & 1.25 - 1.50 & 1.008 & $\pm$0.066 & $\pm$0.069 & $\pm$0.09 \\ 
\bacy & 1.50 - 1.75 & 1.000 & $\pm$0.061 & $\pm$0.069 & $\pm$0.089 \\ 
\bacy & 1.75 - 2.00 & 1.058 & $\pm$0.066 & $\pm$0.073 & $\pm$0.095 \\ 
\bacy & 2.00 - 2.25 & 1.058 & $\pm$0.062 & $\pm$0.073 & $\pm$0.095 \\ 
\bacy & 2.25 - 2.50 & 1.252 & $\pm$0.072 & $\pm$0.086 & $\pm$0.112 \\ 
\bacy & 2.50 - 2.75 & 1.240 & $\pm$0.076 & $\pm$0.085 & $\pm$0.111 \\ 
\bacy & 2.75 - 3.00 & 1.249 & $\pm$0.083 & $\pm$0.086 & $\pm$0.112 \\ 
\bacy & 3.00 - 3.25 & 1.230 & $\pm$0.095 & $\pm$0.085 & $\pm$0.11 \\ 
\bacy & 3.25 - 3.50 & 1.088 & $\pm$0.093 & $\pm$0.075 & $\pm$0.097 \\ 
\bacy & 3.50 - 3.75 & 1.406 & $\pm$0.13 & $\pm$0.097 & $\pm$0.126 \\ 
\bacy & 3.75 - 4.00 & 1.131 & $\pm$0.124 & $\pm$0.078 & $\pm$0.101 \\ 
\bacy & 4.00 - 4.25 & 1.244 & $\pm$0.165 & $\pm$0.086 & $\pm$0.111 \\ 
\bacy & 4.25 - 4.50 & 1.209 & $\pm$0.186 & $\pm$0.084 & $\pm$0.108 \\ 
\bacy & 4.50 - 4.75 & 1.824 & $\pm$0.314 & $\pm$0.127 & $\pm$0.163 \\ 
\bacy & 4.75 - 5.00 & 1.508 & $\pm$0.329 & $\pm$0.106 & $\pm$0.135 \\ 
\bacy & 5.00 - 5.50 & 1.182 & $\pm$0.33 & $\pm$0.082 & $\pm$0.106 \\ 
\bacy & 5.50 - 6.00 & 1.171 & $\pm$0.522 & $\pm$0.082 & $\pm$0.105 \\ 
\bacy & 6.00 - 7.00 & 1.848 & $\pm$0.83 & $\pm$0.128 & $\pm$0.165 \\ 
\bacy & 7.00 - 8.00 & 2.079 & $\pm$1.364 & $\pm$0.146 & $\pm$0.186 \\ 
\\
\midy & 0.0 - 0.5 & 0.69 & $\pm$0.071 & $\pm$0.094 & $\pm$0.059 \\ 
\midy & 0.5 - 1.0 & 0.60 & $\pm$0.04 & $\pm$0.081 & $\pm$0.051 \\ 
\midy & 1.0 - 1.5 & 0.65 & $\pm$0.042 & $\pm$0.088 & $\pm$0.056 \\ 
\midy & 1.5 - 2.0 & 0.73 & $\pm$0.054 & $\pm$0.097 & $\pm$0.062 \\ 
\midy & 2.0 - 2.5 & 0.88 & $\pm$0.088 & $\pm$0.12 & $\pm$0.075 \\ 
\midy & 2.5 - 2.0 & 0.81 & $\pm$0.11 & $\pm$0.11 & $\pm$0.07 \\ 
\midy & 2.0 - 3.5 & 1.10 & $\pm$0.17 & $\pm$0.15 & $\pm$0.096 \\ 
\midy & 3.5 - 4.0 & 1.10 & $^{+0.25}_{-0.24}$ & $\pm$0.15 & $\pm$0.098 \\ 
\midy & 4.0 - 5.0 & 1.10 & $^{+0.26}_{-0.24}$ & $\pm$0.15 & $\pm$0.095 \\ 
\midy & 5.0 - 7.0 & 0.59 & $^{+0.21}_{-0.19}$ & $^{+0.073}_{-0.087}$ & $\pm$0.051 \\ 
\midy & 7.0 - 9.0 & 1.10 & $^{+0.61}_{-0.54}$ & $^{+0.13}_{-0.17}$ & $\pm$0.095 \\ 
\\
\fory & 0.00 - 0.25 & 0.566 & $\pm$0.054 & $\pm$0.042 & $\pm$0.051 \\ 
\fory & 0.25 - 0.50 & 0.557 & $\pm$0.046 & $\pm$0.042 & $\pm$0.05 \\ 
\fory & 0.50 - 0.75 & 0.557 & $\pm$0.034 & $\pm$0.042 & $\pm$0.05 \\ 
\fory & 0.75 - 1.00 & 0.563 & $\pm$0.031 & $\pm$0.042 & $\pm$0.05 \\ 
\fory & 1.00 - 1.25 & 0.559 & $\pm$0.029 & $\pm$0.042 & $\pm$0.05 \\ 
\fory & 1.25 - 1.50 & 0.594 & $\pm$0.029 & $\pm$0.044 & $\pm$0.053 \\ 
\fory & 1.50 - 1.75 & 0.634 & $\pm$0.031 & $\pm$0.047 & $\pm$0.057 \\ 
\fory & 1.75 - 2.00 & 0.587 & $\pm$0.029 & $\pm$0.044 & $\pm$0.052 \\ 
\fory & 2.00 - 2.25 & 0.698 & $\pm$0.034 & $\pm$0.052 & $\pm$0.062 \\ 
\fory & 2.25 - 2.50 & 0.691 & $\pm$0.037 & $\pm$0.051 & $\pm$0.062 \\ 
\fory & 2.50 - 2.75 & 0.691 & $\pm$0.041 & $\pm$0.052 & $\pm$0.062 \\ 
\fory & 2.75 - 3.00 & 0.713 & $\pm$0.047 & $\pm$0.053 & $\pm$0.064 \\ 
\fory & 3.00 - 3.25 & 0.812 & $\pm$0.056 & $\pm$0.061 & $\pm$0.073 \\ 
\fory & 3.25 - 3.50 & 0.631 & $\pm$0.052 & $\pm$0.047 & $\pm$0.056 \\ 
\fory & 3.50 - 3.75 & 0.708 & $\pm$0.064 & $\pm$0.053 & $\pm$0.063 \\ 
\fory & 3.75 - 4.00 & 0.772 & $\pm$0.081 & $\pm$0.058 & $\pm$0.069 \\ 
\fory & 4.00 - 4.25 & 0.922 & $\pm$0.112 & $\pm$0.069 & $\pm$0.083 \\ 
\fory & 4.25 - 4.50 & 0.834 & $\pm$0.124 & $\pm$0.062 & $\pm$0.075 \\ 
\fory & 4.50 - 4.75 & 0.953 & $\pm$0.163 & $\pm$0.071 & $\pm$0.085 \\ 
\fory & 4.75 - 5.00 & 0.922 & $\pm$0.183 & $\pm$0.07 & $\pm$0.082 \\ 
\fory & 5.00 - 5.50 & 0.735 & $\pm$0.192 & $\pm$0.055 & $\pm$0.066 \\ 
\fory & 5.50 - 6.00 & 0.997 & $\pm$0.396 & $\pm$0.075 & $\pm$0.089 \\ 
\fory & 6.00 - 7.00 & 0.971 & $\pm$0.378 & $\pm$0.073 & $\pm$0.087 \\ 
\fory & 7.00 - 8.00 & 0.989 & $\pm$0.739 & $\pm$0.075 & $\pm$0.089 \\ 
\end{tabular}\end{ruledtabular}
\end{table}

%====================================================== Table_XI
\begin{table}[!ht]
\caption{\label{tab:rdaucent1}
Data tables for \rdau as a function of \pt for 20--40\% centrality.}
%\tiny
\begin{ruledtabular}\begin{tabular}{cccccc}
y & \pt [\gevc] & \rdau & Type A & Type B & Type C \\
\hline
\bacy & 0.00 - 0.25 & 0.706 & $\pm$0.123 & $\pm$0.049 & $\pm$0.06 \\ 
\bacy & 0.25 - 0.50 & 0.739 & $\pm$0.108 & $\pm$0.052 & $\pm$0.063 \\ 
\bacy & 0.50 - 0.75 & 0.815 & $\pm$0.113 & $\pm$0.057 & $\pm$0.07 \\ 
\bacy & 0.75 - 1.00 & 0.925 & $\pm$0.069 & $\pm$0.064 & $\pm$0.079 \\ 
\bacy & 1.00 - 1.25 & 0.984 & $\pm$0.051 & $\pm$0.068 & $\pm$0.084 \\ 
\bacy & 1.25 - 1.50 & 0.874 & $\pm$0.059 & $\pm$0.06 & $\pm$0.075 \\ 
\bacy & 1.50 - 1.75 & 1.021 & $\pm$0.054 & $\pm$0.07 & $\pm$0.087 \\ 
\bacy & 1.75 - 2.00 & 1.008 & $\pm$0.054 & $\pm$0.069 & $\pm$0.086 \\ 
\bacy & 2.00 - 2.25 & 1.067 & $\pm$0.06 & $\pm$0.073 & $\pm$0.091 \\ 
\bacy & 2.25 - 2.50 & 1.169 & $\pm$0.072 & $\pm$0.08 & $\pm$0.1 \\ 
\bacy & 2.50 - 2.75 & 1.103 & $\pm$0.073 & $\pm$0.076 & $\pm$0.094 \\ 
\bacy & 2.75 - 3.00 & 1.137 & $\pm$0.084 & $\pm$0.078 & $\pm$0.097 \\ 
\bacy & 3.00 - 3.25 & 1.160 & $\pm$0.101 & $\pm$0.08 & $\pm$0.099 \\ 
\bacy & 3.25 - 3.50 & 1.099 & $\pm$0.102 & $\pm$0.076 & $\pm$0.094 \\ 
\bacy & 3.50 - 3.75 & 1.234 & $\pm$0.131 & $\pm$0.085 & $\pm$0.106 \\ 
\bacy & 3.75 - 4.00 & 0.957 & $\pm$0.121 & $\pm$0.066 & $\pm$0.082 \\ 
\bacy & 4.00 - 4.25 & 1.376 & $\pm$0.195 & $\pm$0.095 & $\pm$0.118 \\ 
\bacy & 4.25 - 4.50 & 1.077 & $\pm$0.19 & $\pm$0.075 & $\pm$0.092 \\ 
\bacy & 4.50 - 4.75 & 1.401 & $\pm$0.28 & $\pm$0.097 & $\pm$0.12 \\ 
\bacy & 4.75 - 5.00 & 1.028 & $\pm$0.266 & $\pm$0.072 & $\pm$0.088 \\ 
\bacy & 5.00 - 5.50 & 1.481 & $\pm$0.496 & $\pm$0.103 & $\pm$0.127 \\ 
\bacy & 5.50 - 6.00 & 1.188 & $\pm$0.641 & $\pm$0.083 & $\pm$0.102 \\ 
\bacy & 6.00 - 7.00 & 0.793 & $\pm$0.469 & $\pm$0.055 & $\pm$0.068 \\ 
\bacy & 7.00 - 8.00 & 0.996 & $\pm$0.836 & $\pm$0.07 & $\pm$0.085 \\ 
\\
\midy & 0.0 - 0.5 & 0.83 & $\pm$0.089 & $\pm$0.11 & $\pm$0.068 \\ 
\midy & 0.5 - 1.0 & 0.80 & $\pm$0.055 & $\pm$0.11 & $\pm$0.065 \\ 
\midy & 1.0 - 1.5 & 0.69 & $\pm$0.05 & $\pm$0.093 & $\pm$0.056 \\ 
\midy & 1.5 - 2.0 & 0.85 & $\pm$0.069 & $\pm$0.11 & $\pm$0.069 \\ 
\midy & 2.0 - 2.5 & 0.97 & $\pm$0.11 & $\pm$0.13 & $\pm$0.079 \\ 
\midy & 2.5 - 2.0 & 0.82 & $\pm$0.12 & $\pm$0.11 & $\pm$0.067 \\ 
\midy & 2.0 - 3.5 & 0.97 & $\pm$0.18 & $\pm$0.13 & $\pm$0.079 \\ 
\midy & 3.5 - 4.0 & 1.40 & $\pm$0.31 & $\pm$0.18 & $\pm$0.11 \\ 
\midy & 4.0 - 5.0 & 1.20 & $^{+0.3}_{-0.28}$ & $\pm$0.16 & $\pm$0.096 \\ 
\midy & 5.0 - 7.0 & 1.20 & $^{+0.33}_{-0.32}$ & $^{+0.15}_{-0.18}$ & $\pm$0.098 \\ 
\midy & 7.0 - 9.0 & 0.47 & $^{+0.43}_{-0.25}$ & $^{+0.054}_{-0.071}$ & $\pm$0.038 \\ 
\\
\fory & 0.00 - 0.25 & 0.649 & $\pm$0.076 & $\pm$0.049 & $\pm$0.056 \\ 
\fory & 0.25 - 0.50 & 0.735 & $\pm$0.066 & $\pm$0.056 & $\pm$0.063 \\ 
\fory & 0.50 - 0.75 & 0.680 & $\pm$0.065 & $\pm$0.051 & $\pm$0.058 \\ 
\fory & 0.75 - 1.00 & 0.697 & $\pm$0.05 & $\pm$0.052 & $\pm$0.06 \\ 
\fory & 1.00 - 1.25 & 0.661 & $\pm$0.042 & $\pm$0.049 & $\pm$0.057 \\ 
\fory & 1.25 - 1.50 & 0.653 & $\pm$0.035 & $\pm$0.048 & $\pm$0.056 \\ 
\fory & 1.50 - 1.75 & 0.712 & $\pm$0.037 & $\pm$0.053 & $\pm$0.061 \\ 
\fory & 1.75 - 2.00 & 0.701 & $\pm$0.037 & $\pm$0.052 & $\pm$0.06 \\ 
\fory & 2.00 - 2.25 & 0.716 & $\pm$0.039 & $\pm$0.053 & $\pm$0.061 \\ 
\fory & 2.25 - 2.50 & 0.818 & $\pm$0.047 & $\pm$0.061 & $\pm$0.07 \\ 
\fory & 2.50 - 2.75 & 0.831 & $\pm$0.052 & $\pm$0.062 & $\pm$0.071 \\ 
\fory & 2.75 - 3.00 & 0.831 & $\pm$0.057 & $\pm$0.062 & $\pm$0.071 \\ 
\fory & 3.00 - 3.25 & 0.844 & $\pm$0.065 & $\pm$0.063 & $\pm$0.072 \\ 
\fory & 3.25 - 3.50 & 0.728 & $\pm$0.066 & $\pm$0.054 & $\pm$0.062 \\ 
\fory & 3.50 - 3.75 & 0.752 & $\pm$0.075 & $\pm$0.056 & $\pm$0.064 \\ 
\fory & 3.75 - 4.00 & 0.819 & $\pm$0.095 & $\pm$0.061 & $\pm$0.07 \\ 
\fory & 4.00 - 4.25 & 1.191 & $\pm$0.148 & $\pm$0.089 & $\pm$0.102 \\ 
\fory & 4.25 - 4.50 & 1.252 & $\pm$0.192 & $\pm$0.094 & $\pm$0.107 \\ 
\fory & 4.50 - 4.75 & 1.276 & $\pm$0.223 & $\pm$0.096 & $\pm$0.109 \\ 
\fory & 4.75 - 5.00 & 1.206 & $\pm$0.247 & $\pm$0.091 & $\pm$0.103 \\ 
\fory & 5.00 - 5.50 & 0.859 & $\pm$0.237 & $\pm$0.065 & $\pm$0.074 \\ 
\fory & 5.50 - 6.00 & 0.960 & $\pm$0.356 & $\pm$0.072 & $\pm$0.082 \\ 
\fory & 6.00 - 7.00 & 1.111 & $\pm$0.448 & $\pm$0.083 & $\pm$0.095 \\ 
\fory & 7.00 - 8.00 & 1.519 & $\pm$1.038 & $\pm$0.115 & $\pm$0.13 \\ 
\end{tabular}\end{ruledtabular}
\end{table}

%====================================================== Table_XII
\begin{table}[!ht]
\caption{\label{tab:rdaucent2}
 Data tables for \rdau as a function of \pt for 40--60\% centrality.}
\centering
%\tiny
\begin{ruledtabular}\begin{tabular}{cccccc}
y & \pt [\gevc] & \rdau & Type A & Type B & Type C \\
\hline
\bacy & 0.00 - 0.25 & 0.953 & $\pm$0.103 & $\pm$0.066 & $\pm$0.087 \\ 
\bacy & 0.25 - 0.50 & 0.865 & $\pm$0.104 & $\pm$0.061 & $\pm$0.079 \\ 
\bacy & 0.50 - 0.75 & 0.995 & $\pm$0.094 & $\pm$0.069 & $\pm$0.09 \\ 
\bacy & 0.75 - 1.00 & 0.920 & $\pm$0.06 & $\pm$0.063 & $\pm$0.084 \\ 
\bacy & 1.00 - 1.25 & 0.976 & $\pm$0.055 & $\pm$0.067 & $\pm$0.089 \\ 
\bacy & 1.25 - 1.50 & 0.929 & $\pm$0.056 & $\pm$0.064 & $\pm$0.084 \\ 
\bacy & 1.50 - 1.75 & 0.979 & $\pm$0.067 & $\pm$0.067 & $\pm$0.089 \\ 
\bacy & 1.75 - 2.00 & 1.018 & $\pm$0.06 & $\pm$0.07 & $\pm$0.093 \\ 
\bacy & 2.00 - 2.25 & 1.015 & $\pm$0.074 & $\pm$0.07 & $\pm$0.092 \\ 
\bacy & 2.25 - 2.50 & 1.159 & $\pm$0.091 & $\pm$0.08 & $\pm$0.105 \\ 
\bacy & 2.50 - 2.75 & 1.227 & $\pm$0.086 & $\pm$0.085 & $\pm$0.112 \\ 
\bacy & 2.75 - 3.00 & 1.036 & $\pm$0.088 & $\pm$0.071 & $\pm$0.094 \\ 
\bacy & 3.00 - 3.25 & 1.046 & $\pm$0.101 & $\pm$0.072 & $\pm$0.095 \\ 
\bacy & 3.25 - 3.50 & 1.114 & $\pm$0.114 & $\pm$0.077 & $\pm$0.101 \\ 
\bacy & 3.50 - 3.75 & 1.152 & $\pm$0.14 & $\pm$0.08 & $\pm$0.105 \\ 
\bacy & 3.75 - 4.00 & 1.124 & $\pm$0.153 & $\pm$0.078 & $\pm$0.102 \\ 
\bacy & 4.00 - 4.25 & 1.313 & $\pm$0.212 & $\pm$0.091 & $\pm$0.119 \\ 
\bacy & 4.25 - 4.50 & 1.427 & $\pm$0.253 & $\pm$0.099 & $\pm$0.13 \\ 
\bacy & 4.50 - 4.75 & 1.080 & $\pm$0.263 & $\pm$0.075 & $\pm$0.098 \\ 
\bacy & 4.75 - 5.00 & 1.365 & $\pm$0.356 & $\pm$0.096 & $\pm$0.124 \\ 
\bacy & 5.00 - 5.50 & 0.985 & $\pm$0.345 & $\pm$0.069 & $\pm$0.09 \\ 
\bacy & 5.50 - 6.00 & 1.354 & $\pm$0.796 & $\pm$0.094 & $\pm$0.123 \\ 
\bacy & 6.00 - 7.00 & 1.297 & $\pm$0.671 & $\pm$0.09 & $\pm$0.118 \\ 
\bacy & 7.00 - 8.00 & 4.130 & $\pm$2.905 & $\pm$0.291 & $\pm$0.375 \\ 
\\
\midy & 0.0 - 0.5 & 0.94 & $\pm$0.11 & $\pm$0.13 & $\pm$0.082 \\ 
\midy & 0.5 - 1.0 & 0.73 & $\pm$0.059 & $\pm$0.098 & $\pm$0.063 \\ 
\midy & 1.0 - 1.5 & 0.73 & $\pm$0.059 & $\pm$0.099 & $\pm$0.064 \\ 
\midy & 1.5 - 2.0 & 0.80 & $\pm$0.074 & $\pm$0.11 & $\pm$0.07 \\ 
\midy & 2.0 - 2.5 & 0.99 & $\pm$0.12 & $\pm$0.14 & $\pm$0.087 \\ 
\midy & 2.5 - 2.0 & 0.87 & $\pm$0.15 & $\pm$0.12 & $\pm$0.076 \\ 
\midy & 2.0 - 3.5 & 0.82 & $^{+0.2}_{-0.19}$ & $\pm$0.11 & $\pm$0.072 \\ 
\midy & 3.5 - 4.0 & 0.73 & $^{+0.29}_{-0.26}$ & $\pm$0.098 & $\pm$0.064 \\ 
\midy & 4.0 - 5.0 & 1.10 & $^{+0.34}_{-0.3}$ & $\pm$0.14 & $\pm$0.094 \\ 
\midy & 5.0 - 7.0 & 1.60 & $^{+0.47}_{-0.45}$ & $^{+0.2}_{-0.23}$ & $\pm$0.14 \\ 
\midy & 7.0 - 9.0 & 0.10 & $^{+0.67}_{-0.51}$ & $\pm$0.012 & $\pm$0.009 \\ 
\\
\fory & 0.00 - 0.25 & 0.897 & $\pm$0.098 & $\pm$0.067 & $\pm$0.082 \\ 
\fory & 0.25 - 0.50 & 0.731 & $\pm$0.087 & $\pm$0.055 & $\pm$0.067 \\ 
\fory & 0.50 - 0.75 & 0.816 & $\pm$0.093 & $\pm$0.061 & $\pm$0.074 \\ 
\fory & 0.75 - 1.00 & 0.778 & $\pm$0.046 & $\pm$0.058 & $\pm$0.071 \\ 
\fory & 1.00 - 1.25 & 0.802 & $\pm$0.042 & $\pm$0.06 & $\pm$0.073 \\ 
\fory & 1.25 - 1.50 & 0.862 & $\pm$0.046 & $\pm$0.064 & $\pm$0.078 \\ 
\fory & 1.50 - 1.75 & 0.916 & $\pm$0.045 & $\pm$0.068 & $\pm$0.083 \\ 
\fory & 1.75 - 2.00 & 0.865 & $\pm$0.047 & $\pm$0.064 & $\pm$0.079 \\ 
\fory & 2.00 - 2.25 & 0.859 & $\pm$0.051 & $\pm$0.064 & $\pm$0.078 \\ 
\fory & 2.25 - 2.50 & 0.854 & $\pm$0.056 & $\pm$0.064 & $\pm$0.078 \\ 
\fory & 2.50 - 2.75 & 0.781 & $\pm$0.057 & $\pm$0.058 & $\pm$0.071 \\ 
\fory & 2.75 - 3.00 & 0.933 & $\pm$0.072 & $\pm$0.069 & $\pm$0.085 \\ 
\fory & 3.00 - 3.25 & 0.907 & $\pm$0.079 & $\pm$0.068 & $\pm$0.082 \\ 
\fory & 3.25 - 3.50 & 0.807 & $\pm$0.081 & $\pm$0.06 & $\pm$0.073 \\ 
\fory & 3.50 - 3.75 & 0.878 & $\pm$0.096 & $\pm$0.066 & $\pm$0.08 \\ 
\fory & 3.75 - 4.00 & 0.868 & $\pm$0.115 & $\pm$0.065 & $\pm$0.079 \\ 
\fory & 4.00 - 4.25 & 1.109 & $\pm$0.161 & $\pm$0.083 & $\pm$0.101 \\ 
\fory & 4.25 - 4.50 & 0.911 & $\pm$0.167 & $\pm$0.068 & $\pm$0.083 \\ 
\fory & 4.50 - 4.75 & 1.082 & $\pm$0.23 & $\pm$0.081 & $\pm$0.098 \\ 
\fory & 4.75 - 5.00 & 1.246 & $\pm$0.281 & $\pm$0.094 & $\pm$0.113 \\ 
\fory & 5.00 - 5.50 & 1.114 & $\pm$0.343 & $\pm$0.084 & $\pm$0.101 \\ 
\fory & 5.50 - 6.00 & 1.417 & $\pm$0.632 & $\pm$0.107 & $\pm$0.129 \\ 
\fory & 6.00 - 7.00 & 1.268 & $\pm$0.55 & $\pm$0.095 & $\pm$0.115 \\ 
\fory & 7.00 - 8.00 & 0.747 & $\pm$7.473e+39 & $\pm$0.057 & $\pm$0.068 \\ 
\end{tabular}\end{ruledtabular}
\end{table}

%====================================================== Table_XIII
\begin{table}[!ht]
\caption{\label{tab:rdaucent3}
 Data tables for \rdau as a function of \pt for 60--88\% centrality.}
\centering
%\tiny
\begin{ruledtabular}\begin{tabular}{cccccc}
y & \pt [\gevc] & \rdau & Type A & Type B & Type C \\
\hline
\bacy & 0.00 - 0.25 & 0.829 & $\pm$0.14 & $\pm$0.057 & $\pm$0.088 \\ 
\bacy & 0.25 - 0.50 & 0.823 & $\pm$0.091 & $\pm$0.058 & $\pm$0.088 \\ 
\bacy & 0.50 - 0.75 & 0.922 & $\pm$0.074 & $\pm$0.064 & $\pm$0.098 \\ 
\bacy & 0.75 - 1.00 & 0.965 & $\pm$0.068 & $\pm$0.066 & $\pm$0.103 \\ 
\bacy & 1.00 - 1.25 & 0.947 & $\pm$0.057 & $\pm$0.065 & $\pm$0.101 \\ 
\bacy & 1.25 - 1.50 & 0.965 & $\pm$0.059 & $\pm$0.066 & $\pm$0.103 \\ 
\bacy & 1.50 - 1.75 & 0.990 & $\pm$0.058 & $\pm$0.068 & $\pm$0.105 \\ 
\bacy & 1.75 - 2.00 & 0.888 & $\pm$0.059 & $\pm$0.061 & $\pm$0.095 \\ 
\bacy & 2.00 - 2.25 & 1.097 & $\pm$0.077 & $\pm$0.075 & $\pm$0.117 \\ 
\bacy & 2.25 - 2.50 & 1.073 & $\pm$0.084 & $\pm$0.074 & $\pm$0.114 \\ 
\bacy & 2.50 - 2.75 & 1.047 & $\pm$0.09 & $\pm$0.072 & $\pm$0.111 \\ 
\bacy & 2.75 - 3.00 & 1.175 & $\pm$0.107 & $\pm$0.081 & $\pm$0.125 \\ 
\bacy & 3.00 - 3.25 & 1.132 & $\pm$0.122 & $\pm$0.078 & $\pm$0.12 \\ 
\bacy & 3.25 - 3.50 & 0.908 & $\pm$0.115 & $\pm$0.063 & $\pm$0.097 \\ 
\bacy & 3.50 - 3.75 & 0.926 & $\pm$0.138 & $\pm$0.064 & $\pm$0.099 \\ 
\bacy & 3.75 - 4.00 & 0.804 & $\pm$0.136 & $\pm$0.056 & $\pm$0.086 \\ 
\bacy & 4.00 - 4.25 & 1.277 & $\pm$0.234 & $\pm$0.088 & $\pm$0.136 \\ 
\bacy & 4.25 - 4.50 & 0.642 & $\pm$0.187 & $\pm$0.044 & $\pm$0.068 \\ 
\bacy & 4.50 - 4.75 & 1.827 & $\pm$0.401 & $\pm$0.127 & $\pm$0.194 \\ 
\bacy & 4.75 - 5.00 & 0.700 & $\pm$0.251 & $\pm$0.049 & $\pm$0.075 \\ 
\bacy & 5.00 - 5.50 & 1.193 & $\pm$0.42 & $\pm$0.083 & $\pm$0.127 \\ 
\bacy & 5.50 - 6.00 & 3.141 & $\pm$1.737 & $\pm$0.219 & $\pm$0.334 \\ 
\bacy & 6.00 - 7.00 & 1.122 & $\pm$0.714 & $\pm$0.078 & $\pm$0.119 \\ 
\bacy & 7.00 - 8.00 & 0.443 & $\pm$4.427e+39 & $\pm$0.031 & $\pm$0.047 \\ 
\\
\midy & 0.0 - 0.5 & 0.91 & $\pm$0.13 & $\pm$0.12 & $\pm$0.094 \\ 
\midy & 0.5 - 1.0 & 0.88 & $\pm$0.077 & $\pm$0.12 & $\pm$0.091 \\ 
\midy & 1.0 - 1.5 & 0.77 & $\pm$0.069 & $\pm$0.1 & $\pm$0.08 \\ 
\midy & 1.5 - 2.0 & 0.86 & $\pm$0.089 & $\pm$0.11 & $\pm$0.089 \\ 
\midy & 2.0 - 2.5 & 0.87 & $\pm$0.13 & $\pm$0.12 & $\pm$0.09 \\ 
\midy & 2.5 - 2.0 & 1.10 & $\pm$0.18 & $\pm$0.14 & $\pm$0.11 \\ 
\midy & 2.0 - 3.5 & 1.00 & $\pm$0.24 & $\pm$0.14 & $\pm$0.11 \\ 
\midy & 3.5 - 4.0 & 1.00 & $^{+0.37}_{-0.35}$ & $\pm$0.14 & $\pm$0.11 \\ 
\midy & 4.0 - 5.0 & 0.81 & $^{+0.34}_{-0.27}$ & $\pm$0.11 & $\pm$0.084 \\ 
\midy & 5.0 - 7.0 & 0.59 & $^{+0.34}_{-0.27}$ & $^{+0.073}_{-0.086}$ & $\pm$0.061 \\ 
\midy & 7.0 - 9.0 & 1.30 & $^{+0.94}_{-0.75}$ & $^{+0.15}_{-0.2}$ & $\pm$0.14 \\ 
\\
\fory & 0.00 - 0.25 & 1.038 & $\pm$0.123 & $\pm$0.078 & $\pm$0.111 \\ 
\fory & 0.25 - 0.50 & 1.100 & $\pm$0.089 & $\pm$0.083 & $\pm$0.117 \\ 
\fory & 0.50 - 0.75 & 0.961 & $\pm$0.064 & $\pm$0.072 & $\pm$0.102 \\ 
\fory & 0.75 - 1.00 & 0.873 & $\pm$0.065 & $\pm$0.065 & $\pm$0.093 \\ 
\fory & 1.00 - 1.25 & 0.859 & $\pm$0.05 & $\pm$0.064 & $\pm$0.091 \\ 
\fory & 1.25 - 1.50 & 0.800 & $\pm$0.051 & $\pm$0.059 & $\pm$0.085 \\ 
\fory & 1.50 - 1.75 & 1.028 & $\pm$0.054 & $\pm$0.076 & $\pm$0.109 \\ 
\fory & 1.75 - 2.00 & 0.963 & $\pm$0.058 & $\pm$0.072 & $\pm$0.103 \\ 
\fory & 2.00 - 2.25 & 0.864 & $\pm$0.062 & $\pm$0.064 & $\pm$0.092 \\ 
\fory & 2.25 - 2.50 & 0.939 & $\pm$0.068 & $\pm$0.07 & $\pm$0.1 \\ 
\fory & 2.50 - 2.75 & 0.966 & $\pm$0.076 & $\pm$0.072 & $\pm$0.103 \\ 
\fory & 2.75 - 3.00 & 1.146 & $\pm$0.094 & $\pm$0.085 & $\pm$0.122 \\ 
\fory & 3.00 - 3.25 & 0.999 & $\pm$0.097 & $\pm$0.075 & $\pm$0.106 \\ 
\fory & 3.25 - 3.50 & 0.886 & $\pm$0.097 & $\pm$0.066 & $\pm$0.094 \\ 
\fory & 3.50 - 3.75 & 0.897 & $\pm$0.112 & $\pm$0.067 & $\pm$0.096 \\ 
\fory & 3.75 - 4.00 & 0.741 & $\pm$0.116 & $\pm$0.055 & $\pm$0.079 \\ 
\fory & 4.00 - 4.25 & 0.985 & $\pm$0.169 & $\pm$0.074 & $\pm$0.105 \\ 
\fory & 4.25 - 4.50 & 1.228 & $\pm$0.227 & $\pm$0.092 & $\pm$0.131 \\ 
\fory & 4.50 - 4.75 & 1.477 & $\pm$0.305 & $\pm$0.111 & $\pm$0.157 \\ 
\fory & 4.75 - 5.00 & 0.906 & $\pm$0.257 & $\pm$0.068 & $\pm$0.096 \\ 
\fory & 5.00 - 5.50 & 0.913 & $\pm$0.309 & $\pm$0.069 & $\pm$0.097 \\ 
\fory & 5.50 - 6.00 & 1.622 & $\pm$0.816 & $\pm$0.122 & $\pm$0.173 \\ 
\fory & 6.00 - 7.00 & 1.381 & $\pm$0.628 & $\pm$0.103 & $\pm$0.147 \\ 
\fory & 7.00 - 8.00 & 1.527 & $\pm$1.12 & $\pm$0.116 & $\pm$0.163 \\ 
\end{tabular}\end{ruledtabular}
\end{table}

\endgroup

\clearpage

%%%%%%%%%%%%%%%%%%%%%%%%%%%  References 

%\bibliography{ppg125x0}   

\end{document}